\documentclass[journal]{IEEEtran}
\usepackage{graphicx}
\usepackage{etoolbox}
\usepackage{multirow}
\usepackage{array}
\usepackage{diagbox}
\usepackage{mathtools}
\usepackage{amsmath}
\usepackage{amssymb}
\usepackage{empheq}
\usepackage{bm}
\usepackage{color}
\usepackage{amsthm}
\usepackage{comment}
\usepackage{algorithm}
\usepackage{algorithmicx}
\usepackage{algpseudocode}
\usepackage{booktabs}
\usepackage{tabularx}
\usepackage{stfloats}
\usepackage{scalefnt}
\usepackage[hidelinks]{hyperref}

\usepackage[caption=false,font=footnotesize]{subfig}

\allowdisplaybreaks
\IEEEoverridecommandlockouts

\def\BibTeX{{\rm B\kern-.05em{\sc i\kern-.025em b}\kern-.08em
    T\kern-.1667em\lower.7ex\hbox{E}\kern-.125emX}}

\IEEEaftertitletext{\vspace{-1\baselineskip}}

\begin{document}
\scalefont{0.99}
\title{Segment-Wise Soft Robotics Inspired Flexible Antenna Arrays: Design and Optimization}

\author{Shuaishuai Han,~\IEEEmembership{Member,~IEEE,} Konstantinos Ntougias, 
\IEEEmembership{Senior Member,~IEEE}, \\ Elio Faddoul, \IEEEmembership{Member,~IEEE}, and Ioannis Krikidis, \IEEEmembership{Fellow,~IEEE}
\vspace{-0.1cm}
\thanks{The authors are with the IRIDA Research Laboratory for Communication Technologies in the Department of Electrical and Computer Engineering, University of Cyprus, Nicosia, Cyprus (e-mail: shan0001@ucy.ac.cy; kntoug01@ucy.ac.cy; efaddo01@ucy.ac.cy; krikidis@ucy.ac.cy).}}

\maketitle

\begin{abstract}
In this paper, we propose a segment-wise soft robotic antenna (SRA) system, where each soft robotic arm referred to as a tentacle, comprises multiple independently controllable segments with bending, elongation–retraction, and sweeping motions. By adjusting segment motion parameters, the positions of surface-mounted antennas are reconfigured, distinguishing it from conventional reconfigurable antenna (RA) systems. Based on this model, we propose two antenna deployment schemes: the segmented end-antenna configuration (SEAC), where fixed antennas are mounted at the segment ends and reconfigured via segment motions; and the hybrid end-and-intermediate antenna configuration (HEIAC), where RAs are further integrated as intra-segment antennas. In HEIAC, soft-robot segment deformation provides large-scale spatial reconfiguration, while RAs enable fine-grained adjustment. For SEAC, we formulate a sum-rate maximization problem accounting for inter-segment connectivity and the nonlinear mapping from segment deformation parameters to antenna coordinates, and develop a penalty dual decomposition–projected gradient ascent (PDD-PGA) algorithm. For HEIAC, we jointly optimize segment deformation, intra-segment antenna positions, and antenna activation using a block coordinate descent (BCD)-PDD-PGA algorithm with greedy backward antenna selection. Simulation results demonstrate that the proposed schemes substantially outperform fixed-position antenna arrays and conventional RA baselines. In particular, SEAC and HEIAC achieve 37.9\% and 32.1\% sum-rate gains over conventional 3D reconfigurable arrays, respectively, while SEAC provides up to a 49.3\% gain in compact array deployments.

\end{abstract}


\section{Introduction}
Over the past few decades, multiple-input multiple-output (MIMO) has become a cornerstone of modern wireless communications, owing to its capability to enhance capacity, reliability, and efficiency. However, fixed antenna geometries fundamentally constrain conventional MIMO performance. In practical propagation environments, wireless channels may vary significantly even with small antenna displacements \cite{fluid_Wong2021}. This location sensitivity indicates that antenna repositioning offers an additional degree of freedom (DoF) to adapt to reshaping array geometry, mitigating spatial correlation and improving user distinguishability \cite{MA_Zhu2026}. This insight has recently sparked interest in reconfigurable antenna-position technologies, such as fluid antenna (FA) systems \cite{fluid_Wong2022}, movable antenna (MA) systems \cite{MA_Zhu2024}, and pinching antenna (PA) systems \cite{pinching_Ding}.

In recent years, soft robots have attracted significant attention as an emerging class of robotic systems distinct from conventional rigid robots, which rely on rigid materials and articulated structures for motion generation \cite{Rus2015}. In contrast, soft robots are continuum systems composed of compliant materials, enabling motion through continuous elastic deformation and smooth body-curvature variations \cite{SRA_Laschi}. Among various soft robotic platforms, soft robotic arms are particularly attractive due to their high shape adaptability and flexible spatial maneuverability. Often inspired by octopus tentacles, these arms can naturally conform to object geometries, making them well suited for tasks like grasping. As illustrated in Fig. 1, a typical manipulation process consists of three successive phases: reaching, grasping, and withdrawing \cite{SRA_Xie}. To realize such continuous deformation and manipulation behaviors, various actuation methods have been developed, among which pneumatic actuation is widely adopted by regulating air pressure inside internal inflatable cavities, such as channels and chambers \cite{SciRep_Elchrif}.

\begin{figure}[t]
	\centering
	\includegraphics[
		width=0.95\columnwidth,
		height=0.3225\textheight,
		keepaspectratio
	]{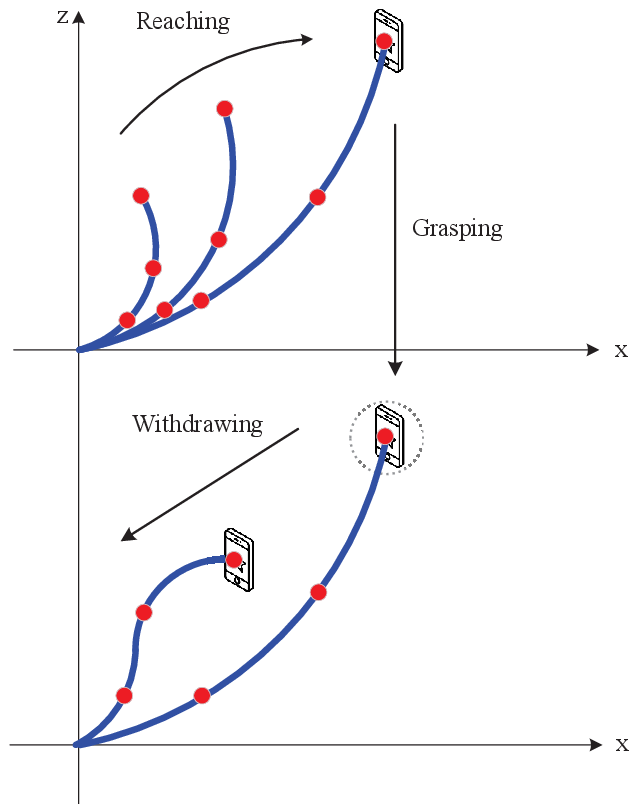}
    \vspace{-0.2cm}
	\caption{Illustration of the manipulation of the soft robotic tentacle.}
	\label{Tentacle motion}
    \vspace{-4mm}
\end{figure}

Owing to the material composition and motion characteristics, soft robots exhibit several distinctive properties, including theoretically infinite DoF, inherent compliance, and safe human--robot interaction capability \cite{TASE_Chen}. Unlike conventional rigid robots, soft robots can deform continuously and are often considered as continuum systems with theoretically infinite DoFs \cite{RA-L_FeliuTalegon,RA-L_Azizkhani}. Moreover, by exploiting compliant materials such as silicone and rubber, soft robots can achieve diverse adaptive and agile motions, including twisting, bending, and grasping \cite{RA-L_Azizkhani}. As a result, they generally offer higher motion flexibility than rigid robots. In addition, their compliant bodies absorb collision energy, enhancing robustness and adaptability to complex environments such as rough terrain. Moreover, materials with compliance comparable to that of biological tissues reduce the risk of unintended harm, making soft robots promising for safe human--robot interaction \cite{Rus2015}. Motivated by these advantages, we incorporate a segment-wise soft robotic architecture into a flexible antenna system, where each tentacle is divided into multiple controllable segments. Antenna elements can be mounted on the segment surfaces \cite{SRA_Xie} and reconfigured through segment deformation. The resulting architecture is referred to as a segment-wise soft-robotic antenna (SRA) array\footnote{Motivated by mechanically feasible segment-wise tentacle actuation in robotics studies, e.g., \cite{SRA_Xie}, we adopt a similar segmentation strategy for communication-oriented antenna repositioning.}.

\subsection{Related Work}
Reconfigurable antenna architectures have recently attracted attention for enhancing wireless performance via adaptive antenna-position and geometry adjustment. In \cite{fluid_Wong2020}, the performance limits of a single-antenna FA system are characterized, where an antenna realized using liquid metals or ionized solutions is assumed to switch instantaneously among multiple ports over a fixed-length linear space. The results show that, even within a limited space, a single-antenna FA system can achieve the capacity of a multi-antenna maximum-ratio combining (MRC) system. In \cite{MA_Ma2024MultiBeam}, multi-beamforming for movable-antenna (MA) arrays, where antenna elements are physically displaced by motors, is investigated. Specifically, the antenna positions and beamforming weights are jointly optimized to enhance the beamforming effectiveness. The MA design is shown to outperform conventional beamforming based on fixed-position antenna (FPA) arrays in terms of beamforming gain and interference suppression. In addition, MA-enabled near-field multiuser communications are studied through movable subarray optimization in \cite{MA_Zhu2025NearField}. In \cite{PASS_Zhou2025}, a sum-rate maximization framework for PA systems is developed, where PAs are activated along dielectric waveguides and their activation positions are jointly optimized with user power allocation. By exploiting position reconfigurability and line-of-sight (LoS)-dominant link enhancement, the PA design achieves substantial performance gains over conventional FPA systems. Furthermore, \cite{PASS_Wang2026} investigates PA systems for indoor immersive communications, accounting for random user interference and PA--user spatial coupling. Unlike conventional reconfigurable antennas based on port switching, motor-driven movement, or waveguide-based activation, this paper deploys antenna elements on soft robotic segment surfaces. This is motivated by recent soft robotic platforms that integrate flexible sensors, processing circuits, and wireless communication modules on the surfaces of tentacle-like soft bodies \cite{SRA_Xie}. The 3D positions of the surface-mounted antennas are then reconfigured through segment deformation.

This above mechanism offers a distinct reconfiguration paradigm with advantages over conventional architectures. To begin, current reconfigurable antenna systems often rely on fixed tracks. For example, in a linear space, such as a fluid container along a mobile-device perimeter, an FA can switch its radiating element among multiple preset ports \cite{fluid_Chai2022}. MAs can be implemented using motorized slides that move antennas along linear \cite{MA_Ma2024Sensing} or circular tracks \cite{Basbug2017}. In PA systems, antennas typically move along waveguides \cite{PASS_Wang2025Beamforming}. Unlike these reconfigurable antennas, SRA systems adjust the 3D positions of surface-mounted antennas without predefined tracks by controlling segment deformation parameters, e.g., bending amplitude and frequency. Recent MA designs enable 2D trackless positioning within a planar region via independent motors \cite{MA_Ma2026ISAC}, or 3D reconfiguration by rotating the entire antenna plane with a rigid robotic arm \cite{MA_Shao2025}. In contrast, rather than relying on a fixed plane, SRA systems reconfigure the antenna array geometry through the 3D deformation of multiple tentacles, thereby providing a richer reconfiguration space and potentially enhanced repositioning flexibility due to their many DoFs. Moreover, existing reconfigurable antennas are typically confined to fixed regions, e.g., a fixed 2D area for MAs \cite{MA_Zhu2025Satellite} or fixed-length linear regions for FAs \cite{fluid_Ghadi2024RIS}. In contrast, SRA systems are mainly limited by physical constraints, such as maximum tentacle bending curvature and twisting angle. Finally, owing to their material compliance, SRA systems are more robust to collisions and better suited for human--robot interaction. Thus, while applicable to conventional urban scenarios, SRA systems are expected to offer greater advantages than traditional reconfigurable antenna systems in complex applications, such as cave exploration \cite{SoftRobot_Gough2021} and human--robot interaction \cite{SoftRobot_Pang2021} for antenna reconfiguration.

\subsection{Motivation and Contributions}
An early investigation of the SRA system was conducted in \cite{FlexibleArray_Faddoul2025}. Specifically, surface-mounted antennas are reconfigured through bending-induced deformation of soft robotic tentacles, demonstrating the feasibility of soft-robot-enabled antenna repositioning. However, the correlation issue is ignored. Specifically, since the antenna elements move collectively with tentacle deformation, the resulting array geometry is strongly coupled. This coupling limits the controllability of individual antenna positions, restricts the achievable spatial DoF for correlation mitigation, and consequently reduces the performance gains enabled by antenna repositioning. Moreover, the nonlinear mapping from tentacle deformation to antenna positions can transform an initially uniform deployment into a highly non-uniform array geometry after bending, thereby significantly increasing antenna correlation. These limitations motivate the segment-wise SRA system proposed in this paper. The main contributions are as follows:

% \footnote{A conference version has been submitted to IEEE GLOBECOM 2026.}:

% Currently, there are too many bullet points. Maybe consider the following:
% one bullet point to say we introduce the mathematical modeling of segment-wise SRAs...
% one bullet point for the SEAC scheme and its optimization approach
% one bullet point for the HEIAC scheme and its approach
% one bullet point for the main results and gains

\begin{itemize}
\item We develop a mathematical model for segment-wise SRAs, in which each tentacle comprises multiple independently controllable segments supporting bending, elongation, and sweeping motions. Elongation enlarges the array aperture, while bending and sweeping enable flexible 3D antenna repositioning. The proposed model also maps segment-wise deformation parameters to antenna coordinates and incorporates inter-segment smoothness constraints to ensure physically feasible deformation. Moreover, we characterize different types of correlation effects inherent in segment-wise SRA systems.

\item Based on the proposed segment-wise SRA model, we propose a segmented end-antenna configuration (SEAC) scheme, where an antenna are placed at each segment endpoint. SEAC reconfigures each antenna's location through physically constrained segment-wise deformation, which differs from conventional reconfigurable antenna systems and helps reduce spatial correlation. We further formulate a sum-rate maximization problem accounting for inter-segment connectivity constraints and the impact of preceding segment deformations on subsequent segment's antenna positions. To solve the resulting geometry-coupled problem, we develop a PDD-PGA algorithm, where the connectivity constraints are embedded into penalty terms for tractable optimization.

\item Furthermore, we propose a hybrid end-and-intermediate antenna configuration (HEIAC) scheme, where reconfigurable antennas, such as MAs, are deployed within the segments of each tentacle, in addition to the terminal antenna. This architecture exploits soft-robot deformation for large-scale antenna reconfiguration while enabling fine-grained intra-segment antenna adjustment. The resulting mixed-integer nonconvex problem jointly involves segment deformation, intra-segment reconfigurable antenna positions, and antenna activation. To solve it efficiently, we develop a block coordinate descent (BCD)-PDD-PGA algorithm, together with a greedy backward search to determine the number of intra-segment antennas.

\item Numerical results demonstrate that the proposed SEAC and HEIAC schemes outperform FPA arrays and conventional 2D/3D reconfigurable antenna baselines, achieving up to a 49.3$\%$ sum-rate gain over the 3D baseline in compact array deployments. The results further reveal the sources of segment-wise gains and the impact of mutual coupling on the achievable performance.
\end{itemize}

\textbf{Notation:} $[\cdot]^T$ and $[\cdot]^H$ denote the transpose and Hermitian transpose, respectively; $\lVert \cdot \rVert_2$ denotes the Euclidean norm

\begin{figure}[ptb]
	\centering
	\includegraphics [scale=0.88]{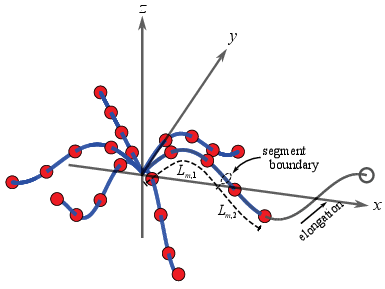}
    \vspace{-0.2cm}
\caption{Illustration of a 3D SRA system with two segments per tentacle.}%
	\label{SRA}
\end{figure}

\section{System Model}
We consider an uplink multi-user single-input multiple-output (MU-SIMO) system, where \(K\) single-antenna user terminals at fixed locations transmit signals to a base station (BS). As shown in Fig.~\ref{SRA}, the BS is equipped with a soft-robot-inspired flexible antenna array comprising \(M\) arms, referred to as tentacles. A segment-wise SRA model is proposed, where each tentacle is partitioned into \(S\) segments, with the \(s\)-th segment comprising \(N_s\) antenna \vspace{0.05cm} elements, \(s\in\{1,\ldots,S\}\), yielding \(N=\sum_{s=1}^{S}N_s\) antenna elements per tentacle. The surface-mounted antenna positions vary with segment deformation. Specifically, each segment is actuated by several parallel silicone bellows, with segment deformation enabled by local pneumatic actuation and independent pressure control~\cite{SRA_Xie,SRA_Katzschmann2019}.

\subsection{Segment-wise Soft Robot Antenna Array Model}
\label{system model}

The deformation of each tentacle in 3D space can be characterized by a combination of fundamental motion components:
\begin{enumerate}
    \item Bending: Soft robotic platforms enable curvature variation through bending and provide many DoFs~\cite{RA-L_FeliuTalegon,RA-L_Azizkhani}, thus allowing highly flexible antenna positioning. Without loss of generality, each tentacle is assumed to bend along the \(z\)-axis.
    \item Elongation-Retraction: Material stretching and compression provide an alternative mechanism for antenna repositioning along the tentacle. Existing materials can achieve elongation ratios up to \(870\%\)~\cite{SRA_Tago}, enabling large antenna displacement and spatial reconfigurability.
    \item Sweeping: Horizontal repositioning of the antenna is achieved through a sweeping motion, where the entire tentacle rotates about its base to provide continuous coverage across the horizontal domain.
\end{enumerate}

\begin{figure*}[ptb]
	\centering
	\includegraphics [scale=0.65]{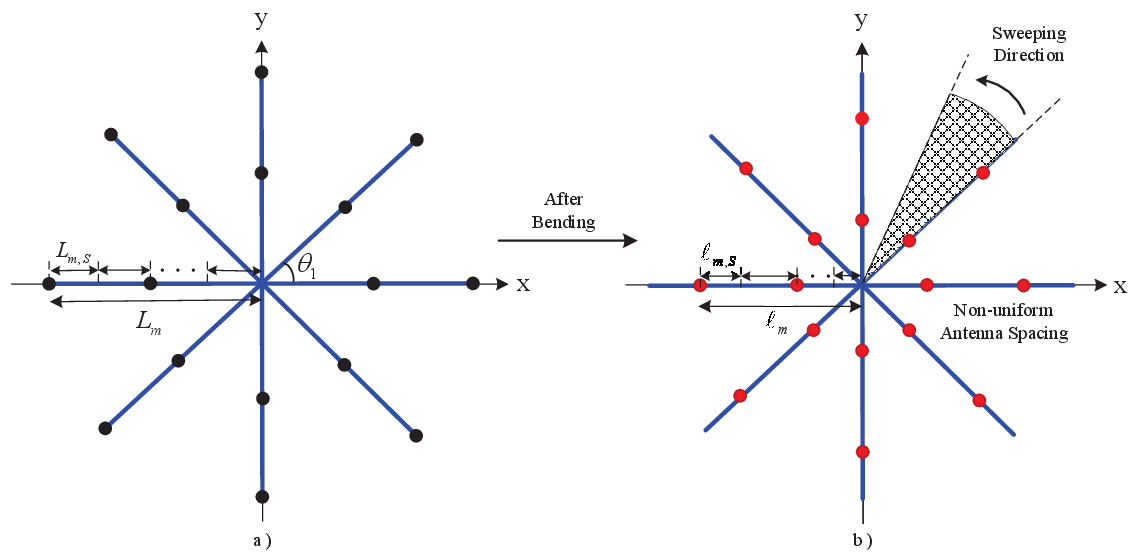 }\vspace{-0.4cm}
\caption{2D projection of antenna locations on segment-wise soft robotic tentacles with two antennas per tentacle as an example. a) Projected antenna locations in the undeformed configuration. b) Projected antenna locations after bending deformation, where the sweeping motion is also illustrated. }%
	\label{Projection}
\end{figure*}

To characterize tentacle deformation, we adopt a sinusoidal profile, which is a communication-oriented reduced-order parametrization rather than a full continuum-mechanics model. It approximates the soft robot’s infinite-dimensional deformation space using a tractable finite-dimensional family of sinusoidal profiles. This choice significantly reduces the optimization complexity while preserving the key geometric variations that affect the antenna positions, inter-element spacing, and array response ~\cite{FlexibleArray_Faddoul2025, TipSR}. Accordingly, the 3D position of an arbitrary point on the \(s\)-th segment of the \(m\)-th tentacle is
\begin{equation}
\label{deformation}
\mathbf r_{m,s}(\ell,t)
=
\big[
\ell\cos\theta_m,\;
\ell\sin\theta_m,\;
z_{m,s}(\ell,t)
\big]^{T}.
\end{equation}
where \(\theta_m\) denotes the azimuth angle of the \(m\)-th tentacle, which is affected by the sweeping motion. Let \(L\) be the arc length from the coordinate origin to an arbitrary point on the \(s\)-th segment of the \(m\)-th tentacle, and let \(\ell\) be its projection length onto the \(xy\)-plane. Due to the bending action along the $z$-axis, the projection length is less than or equal to its actual arc length, i.e., \(\ell \leq L\), as shown in Fig.~\ref{Projection}. Moreover, \(z_{m,s}(\ell,t)\) denotes the vertical sinusoidal displacement induced by the bending motion
\begin{align}
\label{z_ms}z_{m,s}\left(\ell,t\right)=A_{m,s}\sin\!\big(\omega_{m,s} t +v_{m,s}\ell\big),
\end{align}
where the parameter $A_{m,s} \in [0, A_{\max}]$ represents the sinusoidal displacement amplitude, and $v_{m,s} \in [0, v_{\max}]$ denotes the spatial frequency of the segment displacement. The angular frequency $\omega_{m,s}$ and the time variable $t$ are introduced to capture the dynamic bending behavior of the $s$-th segment in the time domain. The SRA geometry is optimized once per channel coherence interval and then kept fixed within the corresponding block. Hence, the explicit dependence on time \(t\) is omitted for notational simplicity.

In practice, the arc length is typically specified, while the corresponding projected length must be determined. Using the Pythagorean relation between the differential planar projection and vertical displacement, the differential arc length is given by
\begin{equation}
\mathrm{d}L
=\sqrt{\mathrm{d}\ell^2+\mathrm{d}z^2}
=\sqrt{1+\left(\frac{\mathrm{d}z}{\mathrm{d}\ell}\right)^2}\,\mathrm{d}\ell,
\end{equation}
where 
\begin{equation}
   \frac{\mathrm{d}z}{\mathrm{d}\ell}
=A_{m,s}v_{m,s}\cos\!\left(v_{m,s}\ell\right). 
\end{equation}

Afterwards, by integrating from $\bar \ell=0$ to $\bar \ell=\ell$, the cumulative arc length of the curve is obtained as
\begin{equation}
\label{projection1}
L(\ell)
=\int_{0}^{\ell}
\sqrt{
1+\left(A_{m,s}v_{m,s}\cos\!\left(v_{m,s}\bar \ell\right)\right)^2
}\,\mathrm{d}\bar \ell,
\end{equation}
where the projected length \(\ell\) is obtained by numerically solving \eqref{projection1} for a given arc length \(L\). 

Due to elongation--retraction motion, the arc length \(L_m\) of the \(m\)-th tentacle is constrained within \([L_{m,\min},L_{m,\max}]\), where \(L_{m,\min}\) and \(L_{m,\max}\) denote its minimum and maximum lengths, respectively. According to \eqref{projection1}, the corresponding projection length varies accordingly. For simplicity, the segment endpoints are uniformly distributed along the tentacle in terms of their cumulative arc lengths. Specifically, the endpoint of the $s$-th segment on the $m$-th tentacle is denoted by $L_{m,s}$, i.e., $L_{m,s}\in[sL_{m,\min}/{S},sL_{m,\max}/{S}]$.

The deformation parameters are governed by the underlying actuation motions. Specifically, elongation--retraction controls the arc length \(L_{m,s}\), sweeping adjusts the azimuth angle \(\theta_m\), and bending determines the sinusoidal parameters \(A_{m,s}\) and \(v_{m,s}\), which characterize the bending magnitude and spatial frequency, respectively. To ensure smooth inter-segment connections and avoid discontinuities or sharp corners in practice, the position function is required to be continuous and differentiable with respect to the actual arc length, following the concept in \cite{SRA_Zhao}. Then, we drive the
continuity constraints $\mathcal{C}^0=0$ and $\mathcal{C}^1=0$, where $\mathcal{C}^0$ and $\mathcal{C}^1$ are compactly written as
\begin{equation}
\label{C^b}
\ \ \ \ \ \ \ \mathcal{C}^b=\left.\frac{\partial^b z_{m,s}(L)}{\partial L^b}\right|_{L=L_{m,s-1}}\!\!\!\!\!\!-
\left.\frac{\partial^b z_{m,s-1}(L)}{\partial L^b}\right|_{L=L_{m,s-1}},
\end{equation}
where $b \in \{0,1\}$ denotes the derivative order. By substituting \eqref{z_ms} into \eqref{C^b}, and invoking the chain rule and the fundamental theorem of calculus, the general expression of $\mathcal{C}^b$ is derived, as shown in \eqref{C_b} at the bottom of the next page.

%\footnote{Strictly speaking, the bending of a soft continuum structure is characterized by variations in its local curvature. To preserve analytical tractability, the bending behavior is equivalently modeled by adjusting the amplitude $A_m$ and the spatial frequency $v_m$ of a sinusoidal deformation, which effectively captures changes in the deformation profile. This abstraction concentrates on the resulting displacement of the antenna elements, which constitutes the primary quantity of interest in the wireless communication analysis.}.

Based on the above formulation, the locations of the antenna elements can be explicitly determined. Thus, the position vector of the \(n\)-th element on the \(s\)-th segment of the \(m\)-th tentacle is
\begin{equation}
\label{3D location}
\mathbf{r}_{m,s,n} = 
\begin{bmatrix}
	x_{m,s,n} \\[1mm]
	y_{m,s,n} \\[1mm]
	z_{m,s,n}
\end{bmatrix} =
\begin{bmatrix}
	\ell(L_{m,s,n})\cos\theta_m  \\[1mm]
	\ell(L_{m,s,n})\sin\theta_m \\[1mm]
	A_{m,s}\sin\!\big(v_{m,s}\ell (L_{m,s,n})\big)
\end{bmatrix}.
\end{equation}
where \(\ell(L_{m,s,n})\) denotes the projected length from the origin to the \(n\)-th antenna element on the \(s\)-th segment of the \(m\)-th tentacle, which can be obtained from \eqref{projection1}. Based on the proposed segment-wise SRA model, we now present the channel model.

\subsection{Channel and Signal Model}
We define $\mathbf{r}_{m,s} \triangleq \left[\mathbf{r}^T_{m,s,1}, \ldots, \mathbf{r}^T_{m,s,N_s}\right]^T \in \mathbb{R}^{3N_s\times 1}$ as the position vector collecting antennas mounted on the $s$-th segment of the $m$-th tentacle. Correspondingly, $\mathbf{r}_m = \left[\mathbf{r}_{m,1}^T, \ldots, \mathbf{r}_{m,S}^T\right]^T \in \mathbb{R}^{3N \times 1}$ denotes the positions of all antennas of the $m$-th tentacle, whereas $\mathbf{r} = \left[\mathbf{r}_1^T, \ldots, \mathbf{r}_M^T\right]^T \in \mathbb{R}^{3MN \times 1}$ denotes positions of all antennas of the SRA array. By using the spatially correlated Rayleigh fading channel model, the channel $\mathbf{h}_{k}\left(\mathbf{r}\right) \in \mathbb{C}^{MN\times 1}$ from the $k$-th user, $k\in\{1,...,K\}$, to all the $MN$ antennas of the SRA array is modeled as \cite{fluid_Espinosa}
\begin{equation} 
\label{channel_k}
\mathbf{h}_{k}\left(\mathbf{r}\right)\sim\mathcal{CN}({0}, \sigma_k^2 {\bf{C}}\left(\mathbf{r}\right)),
\end{equation}
where $\sigma_k^2$ denotes the large-scale channel gain of the $k$th user. The matrix ${\bf C}(\mathbf r)$ captures the spatial correlation among the $MN$ antennas, determined by the proposed antenna configuration scheme in this paper and the antenna positions in \eqref{3D location}. To facilitate analytical derivations, we rewrite the channel model in \eqref{channel_k} with the following equivalent form
\begin{equation}
\mathbf{h}_{k}\left(\mathbf{r}\right)= \sigma_k \, \mathbf{C}^{1/2}(\mathbf{r}) \, \boldsymbol{\eta}_k, 
\end{equation}
where $\mathbf C(\mathbf r)=\mathbf C^{1/2}(\mathbf r)\mathbf C^{1/2}(\mathbf r)$, and $\boldsymbol{\eta}_k\sim\mathcal{CN}(\mathbf 0,\mathbf I_{MN})$ is a complex Gaussian vector. Thus, the channel from all $K$ users to all $MN$ antennas is
$\mathbf H(\mathbf r)=\left[\mathbf h_1(\mathbf r),\ldots,\mathbf h_K(\mathbf r)\right]\in\mathbb C^{MN\times K}$.
After matched filtering, the received signal vector ${\bf y}\in\mathbb C^{MN\times 1}$ at the SRA array is given by
\begin{equation} 
	\label{received signal matrix2}
	{\bf{  y}} = \mathbf{H}\left(\mathbf{r}\right)  {\bf{s}}+ {\bf{  n}}, 
\end{equation}
where $\mathbf{s}\in\mathbb C^{K\times1}$ denotes the $K$-user transmitted signal vector and $\mathbf{n} $ is an additive white Gaussian noise (AWGN) vector. 

\begin{figure*}[!b]
\hrulefill
\begin{align}
\label{C_b}
\mathcal{C}^b=\frac{
A_{m,s} v_{m,s}^{\,b}
\sin\!\left( v_{m,s} \ell(L_{m,s-1}) + \frac{b\pi}{2}\right)
}{
\left[ 1 + \left(A_{m,s} v_{m,s} \cos\!\left(v_{m,s} \ell(L_{m,s-1})\right)\right)^2 \right]^{\frac{b}{2}}
}-
\frac{
A_{m,s-1} v_{m,s-1}^{\,b}
\sin\!\left( v_{m,s-1} \ell(L_{m,s-1}) + \frac{b\pi}{2}\right)
}{
\left[ 1 + \left(A_{m,s-1} v_{m,s-1} \cos\!\left(v_{m,s-1} \ell(L_{m,s-1})\right)\right)^2 \right]^{\frac{b}{2}}
}.
\end{align}
\end{figure*}

To enhance the quality of the received signal, we apply the minimum mean
square error (MMSE) equalizer, thus we obtain 
    \begin{align}
    \label{MMSE}
    {\bf{  x}}={\bf{W}}{\bf{y}}=({\bf{  H}}^H\left(\mathbf{r}\right) {\bf{  H}}\left(\mathbf{r}\right)+1/\gamma{\bf{I}})^{-1}{\bf{ H}}^{H}\left(\mathbf{r}\right)\bf{y},
\end{align}
where ${\bf W}\in\mathbb C^{K\times MN}$ denotes the MMSE matrix, and $\gamma=\sigma_s^2/\sigma_n^2$ represents  the signal-to-noise (SNR) power ratio. We assume perfect channel state information (CSI) at the receiver. The $k$-th entry of the signal vector ${\bf x}\in\mathbb C^{K\times1}$ is
\begin{equation}
x_k 
= \mathbf{w}_k \mathbf{h}_k\left(\mathbf{r}\right) s_k + \sum_{g\in\mathcal{K}\setminus{\{k\}}}\mathbf{w}_k\mathbf{h}_g\left(\mathbf{r}\right)s_g + \mathbf{w}_k{\bf{  n}},
\end{equation}
where $\mathbf{w}_k\in \mathbb{C}^{1 \times MN}$ indicates the MMSE vector corresponding to the $k$-th user and $\mathcal{K}\triangleq\{1,\ldots,K\}$ denotes the user set.

\section{Proposed Antenna Deployment Schemes for Correlation Mitigation}
We exploit segment motions to adjust the 3D coordinates of the surface-mounted antennas, thereby optimizing the array geometry. However, randomly deploying multiple antennas along each tentacle may render the antenna position optimization highly susceptible to spatial correlation issue among antenna elements, which mainly arises from two factors:

\begin{enumerate}
\item For the antennas installed on the same segment or tentacle, any motion (e.g., bending, extension) leads to a global movement of all antennas. This prevents the antennas mounted on the tentacle from moving independently, thereby causing some antennas to be placed in highly unfavorable positions.

\item Due to sinusoidal deformation, uniform spacing in arc length does not generally yield a uniform 3D spatial distribution, owing to the nonlinear mapping from the curve parameter to spatial coordinates. Consequently, certain antenna pairs may exhibit large separations while others are characterized by small inter-element distances.
\end{enumerate}

The above-mentioned factors lead to various correlation effects. To this end, the spatial correlation among antennas on different tentacles is referred to as inter-tentacle correlation, whereas that among antennas on the same tentacle is referred to as intra-tentacle correlation. At the segment level, the intra-tentacle correlation is further divided into inter-segment and intra-segment correlations, which characterize the spatial correlation among antennas on different segments and on the same segment, respectively. The classification of antenna correlation can be found in Table \ref{tab:correlation_classification}.

\begin{table}[t]
\centering
\caption{Classification of Spatial Correlation}
\label{tab:correlation_classification}
\begin{tabularx}{\columnwidth}{@{}
>{\centering\arraybackslash}p{0.24\columnwidth}
>{\centering\arraybackslash}p{0.25\columnwidth}
>{\raggedright\arraybackslash}X
@{}}
\toprule
\textbf{Level} & \textbf{Type} & \textbf{Correlated Antennas} \\
\midrule
Tentacle  & Inter-tentacle & Antennas on different tentacles \\[1mm]
Tentacle  & Intra-tentacle & Antennas on the same tentacle \\[1mm]
Segment & Inter-segment & Antennas on different segments \\[1mm]
Segment & Intra-segment & Antennas on the same segment \\
\bottomrule
\end{tabularx}
\end{table}

\subsection{Segmented End-Antenna Configuration (SEAC)}

To begin, we propose a SEAC scheme, in which each tentacle is divided into multiple segments, with an antenna installed at the end of each segment.  By controlling each segment, we can move the segment-end antenna independently and thus the correlation issue among the antennas on the tentacle can be significantly mitigated. Moreover, the SEAC scheme can benefit from the gains provided by multiple antennas.

To characterize the correlation effects in the SEAC scheme, we introduce the following correlation matrix
\renewcommand{\arraystretch}{1.5}
\begin{equation}
\label{intersegment}
{\bf{C}} \triangleq
\begin{bmatrix}
{\bf{C}}_{1,1}      & {\bf{C}}_{1,2}  & \cdots & {\bf{C}}_{1,M} \\
{\bf{C}}_{2,1} & {\bf{C}}_{2,2}       & \cdots & {\bf{C}}_{2,M} \\
\vdots & \vdots  & \ddots & \vdots \\
{\bf{C}}_{M,1} & {\bf{C}}_{M,2}  & \cdots & {\bf{C}}_{M,M}
\end{bmatrix},
\end{equation}
where the diagonal blocks capture the intra-tentacle correlation, and the off-diagonal blocks characterize the inter-tentacle correlation. The inter-segment correlation arises from both intra-tentacle and inter-tentacle segment correlations. In \eqref{intersegment}, the \((s,q)\)-th entry of \({\bf C}_{m,m'}\) denotes the inter-segment correlation coefficient between the antennas mounted on the \(s\)-th segment of the \(m\)-th tentacle and the \(q\)-th segment of the \(m'\)-th tentacle. Based on the 3D Clarke's model~\cite{fluid_Espinosa}, it is given by
\begin{equation}
c^{(s,q)}_{m,m'} \triangleq \mathrm{sinc}\left(\frac{2\pi}{\lambda}\lVert \mathbf{r}_{m,s} - \mathbf{r}_{m',q} \rVert_2\right),
\end{equation}
where \(\mathrm{sinc}(x)=\sin(x)/x\). Since a single antenna is mounted at the end of each segment, \(\mathbf r_{m,s}\) is obtained from \(\mathbf r_{m,s,n}\) in \eqref{3D location} by omitting the antenna index \(n\).

In the SEAC scheme, the influence of intra-segment correlation is inherently removed, since only a single antenna is installed on each segment, whereas the effects of inter-tentacle and inter-segment correlations can be mitigated through independent control of each segment’s motion.

%However, since the motion of a given segment is inevitably influenced by the motion states of other segments, such correlation is difficult to be completely eliminated.

\subsection{Hybrid End-and-Intermediate Antenna Configuration (HEIAC)}
To further exploit the tentacle capability, we propose a HEIAC scheme, where antennas are mounted at both terminal and intermediate positions of each segment. In HEIAC, the inter-tentacle and inter-segment correlation effects can be mitigated by using the methods in SEAC. However, the intra-segment correlation effect is the most challenging to eliminate, not only due to the global movement of the antennas resulting from segment-level motions, but also due to the close spacing between adjacent antennas mounted on the same segment.

To address this issue, conventional reconfigurable architectures, such as stepper-motor-driven MAs\footnote{In this work, MAs are considered as a representative implementation, while FA systems can also be deployed following mobile-device surface implementations~\cite{fluid_Chai2022}.}, can be integrated on each segment surface as intra-segment antennas\footnote{In tendon-driven soft robotic mechanisms \cite{SRA_Somm2019}, predefined local tracks can be embedded on the surface of each soft segment. Stepper-motor-actuated tendons are routed along these tracks to deform the soft robotic arm and control its shape. In this work, we assume that compact antenna elements are mounted on lightweight carriers and repositioned along the tracks following the same motor-driven tendon actuation principle. This implementation is consistent with the operating principle of MAs, where antenna elements are mechanically displaced within a predefined region, thereby enabling MA-based intra-segment position reconfiguration \cite{MA_Zhu2024}.}.  Although this requires additional actuation and control resources, it can significantly reduce intra-segment correlation. The movement of antennas offers multiple DoFs, benefiting from both the motions of each tentacle and conventional flexible antenna systems. 

In HEIAC, the optimal number of antennas per segment is generally unknown, especially under dynamic segment deformation. When sufficiently many antennas are deployed, adding more elements increases hardware and resource consumption while increasing intra-segment spatial correlation. Hence, HEIAC also optimizes the antenna-number configuration.

To characterize the correlation effects under the HEIAC scheme, we adopt the tentacle-level correlation matrix defined in \eqref{intersegment}. Since HEIAC mounted multiple antennas on each segment, the segment-level correlation matrix $\mathbf{C}_{m,m'}$ used in SEAC need to be modified accordingly, and is given by
\begin{equation}\label{intrasegment}
\mathbf{C}_{m,m'} =
\begin{bmatrix}
\mathbf{C}_{m,m'}^{(1,1)} & \cdots & \mathbf{C}_{m,m'}^{(1,S)} \\
\vdots & \ddots & \vdots \\
\mathbf{C}_{m,m'}^{(S,1)} & \cdots & \mathbf{C}_{m,m'}^{(S,S)}
\end{bmatrix},
\end{equation}
where the block $\mathbf{C}_{m,m'}^{(s,q)}$ with $m = m'$ and $s = q$ represents the intra-segment correlation effect, and all the remaining blocks capture the inter-segment correlation effect.

In \eqref{intrasegment}, the \((i,j)\)-th entry of \(\mathbf{C}_{m,m'}^{(s,q)}\) denotes the correlation coefficient between the \(i\)-th antenna on the \(s\)-th segment of the \(m\)-th tentacle and the \(j\)-th antenna on the \(q\)-th segment of the \(m'\)-th tentacle, and is defined as
\begin{equation}
[\mathbf{C}_{m,m'}^{(s,q)}]_{i,j} \triangleq \mathrm{sinc}\left(\frac{2\pi}{\lambda}\lVert \mathbf{r}_{m,s,i} - \mathbf{r}_{m',q,j} \rVert_2\right),
\end{equation}
where $\mathbf{r}_{m,s,i}$ can be obtained from $\mathbf{r}_{m,s,n}$ in \eqref{3D location}.

\textbf{\textit{Remark 1:}} Compared with SEAC, HEIAC is a distinct scheme that additionally installs the reconfigurable antenna system. The former exploits SRA system's inherent reconfigurability, whereas the latter leverages its compatibility. Their comparison reveals that additional antennas enhance antenna gain but incur higher resource consumption and spatial correlation.

%\
\section{Problem Formation}
The signal-to-interference-plus-noise ratio (SINR) of the $k$-th user can be represented as 
\begin{align}
\label{eq:SINR0}
\gamma_{k}\left(\mathbf{r}\right) = \frac{\left| \mathbf{w}_k \mathbf{h}_k(\mathbf{r}) \right|^2}
{\!\!\sum\limits_{g \in \mathcal{K} \setminus \{k\}\!\!}
\left| \mathbf{w}_k \mathbf{h}_g(\mathbf{r}) \right|^2
\!+ \!\sigma^2||\mathbf{w}_k|| ^2}.
\end{align}

Thus, the achievable rate of the $k$-th user and the sum-rate can be denoted by $R_k\left(\mathbf{r}\right)=\log_2\left(1+\gamma_k\left(\mathbf{r}\right)\right)$ and $R\left(\mathbf{r}\right)=\sum_{k\in\mathcal{K}}R_k\left(\mathbf{r}\right)$, respectively.

In the SEAC scheme, we optimize the motion parameters of each segment on the \(m\)-th tentacle, \(m\in\mathcal M\), under the physical constraints, such that the end-mounted antennas achieve positions that maximize the sum rate. Specifically, the optimization variables include the deformation amplitude \(A_{m,s}\), spatial frequency \(v_{m,s}\), segment length \(L_{m,s}\) of the \(s\)-th segment, and the azimuth angle \(\theta_m\) of the \(m\)-th tentacle. Since each segment has one end-mounted antenna, the antenna index \(n\) is omitted for notational simplicity. The problem is formulated as
\vspace{-1mm}
\begin{subequations}
\begin{align}
\text{(P1):\! }    
\max_{\boldsymbol{\bar \psi}}&
 \sum_{k \in \mathcal{K}}\!
\log_2 \!\!\left(\! 1\!+\!
\frac{\left| \mathbf{w}_k \mathbf{h}_k(\mathbf{r}) \right|^2}
{\!\!\sum\limits_{g \in \mathcal{K} \setminus \{k\}\!\!}
\left| \mathbf{w}_k \mathbf{h}_g(\mathbf{r}) \right|^2
\!+ \!\sigma^2||\mathbf{w}_k|| ^2}
\!\right)  \\
\text{s.t.}\ 
& \  0 \le A_{m,s} \le A_{\max},
 \ \ \ \ \  \ \ \  \ \   \ \ \ \ \ \ \  \ \ \ \ \ \ \! \!    \forall m,s \label{eq:P2b} \\
& \ 0 \le v_{m,s} \le v_{\max}, \ \ \ \ \  \ \ \  \ \ \ \ \ \ \ \ \ \ \ \ \ \ \ \ \ \!
\forall m,s  \label{eq:P2c} \\
& \ \theta_{m-1}+\Delta \theta  \le \theta_m,  \ \  \  \ \  \ \ \ \  \ \ \ \ \ \ \  \ \  \ \ \ \ \  \ \!\forall m\label{eq:P2d}\\
&\  \theta_{m,\min}  \le \theta_m \le \theta_{m,\max}, \ \ \ \ \ \ \ \ \ \ \ \ \  \ \forall m \label{eq:P2e} \\
& \ L_{m,s-1} + \Delta L \le L_{m,s} , \ \  \ \ \ \ \ \ \ \ \  \ \ \ \ \ \ \!  \forall m,s \label{eq:P2f}\\[-0.5mm]
&\  \frac{s}{S}L_{m,\min} \le L_{m,s} \le \frac{s}{S}L_{m,\max}, \  \    \forall m  \label{eq:P2g}\\[-0.5mm]
&\ \mathcal{C}^0=0, \; \mathcal{C}^1=0, \ \ \ \ \ \  \ \ \ \ \ \! \ \ \ \ \ \ \ \   \ \ \ \   \ \ \ \  \ \ \  \forall m,s \label{eq:P2h}
\end{align}
\end{subequations}
where the constraints in \eqref{eq:P2h} ensure smooth continuity at segment boundaries. The constraints \eqref{eq:P2b}, \eqref{eq:P2c}, \eqref{eq:P2e}, and \eqref{eq:P2g} 
define the feasible ranges of $A_m$, $v_m$, $\theta_m$, and $L_{m,s}$. The constraints 
\eqref{eq:P2d} and \eqref{eq:P2f} enforce the monotonic ordering of the azimuth angles 
and segment length $L_{m,s}$, respectively. Unless otherwise specified, $m=1,\ldots,M$ and $s=1,\ldots,S$. 
In constraints involving $\theta_{m-1}$ or $L_{m,s-1}$, the corresponding index ranges are 
$m=2,\ldots,M$ and $s=2,\ldots,S$, respectively. 

The parameter set \(\boldsymbol{\bar \psi}\) \vspace{0.05cm}, collecting the parameters of all \(M\) tentacles, is defined as $\boldsymbol{\bar \psi}
\triangleq
\left\{\boldsymbol{\bar \psi}_m\right\}_{m=1}^{M}
\in \mathbb{R}^{3MS+M}$, where
\begin{equation}
\boldsymbol{\bar \psi}_{m}
\!\triangleq\!
\left[
A_{m,1},\!..., A_{m,S},
v_{m,1},\!..., v_{m,S},
L_{m,1},\!..., L_{m,S},
\theta_{m}
\right]^{ T}\!.
\end{equation}

In the HEIAC scheme, we aim to optimize the antenna positions within each segment with the assistance of MA systems, together with the parameters controlling the motion of each segment, such that each antenna within the $s$-th segment of the $m$-th tentacle, for all $m \in \mathcal{M}$, can reach its desired position. In addition, we aim to balance the achievable rate gain from activating additional antennas against the corresponding resource cost and correlation-induced performance loss. Accordingly, we need to optimize the parameters of the $s$-th segment on the $m$-th tentacle (i.e., $A_{m,s}$, $v_{m,s}$), the antenna position $L_{m,n,s}$, the azimuth angle $\theta_m$ of the $m$-th tentacle, and activated antenna index $a_{m,s,n}$. This problem is formulated as
\vspace{-1mm}
\begin{subequations}
\begin{align}
\text{(P2):\! }  
\max_{\boldsymbol{\tilde \psi}}&
 \sum_{k \in \mathcal{K}}\!
\log_2 \!\!\left(\!1\!+\!
\frac{\left| \mathbf{w}_k \mathbf{h}_k(\mathbf{r}) \right|^2}
{\!\!\sum\limits_{g \in \mathcal{K} \setminus \{k\}\!\!}
\left| \mathbf{w}_k \mathbf{h}_g(\mathbf{r}) \right|^2
\!+ \!\sigma^2||\mathbf{w}_k|| ^2}
\!\right)\nonumber\\&-\eta \sum_{m=1}^{M}\sum_{s=1}^{S}\sum_{n=1}^{N-1} a_{m,s,n}  \\[1mm]
\text{s.t.}\ 
& 0 \le A_{m,s} \le A_{\max}, \ \ \ \ \ \ \ \ \ \ \ \ \ \ \ \ \ \ \ \ \ \ \ \!  \forall m,s
 \\
& 0 \le v_{m,s} \le v_{\max}, \ \ \ \ \ \ \ \ \ \  \ \ \ \ \ \ \ \ \ \  \ \ \ \ \forall m,s  \\
& \theta_{m-1}+\Delta \theta  \le \theta_m, \  \  \ \ \ \ \ \ \   \  \ \ \  \ \  \ \ \  \ \ \ \ \ \  \!\forall m \\
& \theta_{m,\min}  \le \theta_m \le \theta_{m,\max}, \ \ \ \ \ \ \ \  \  \ \ \ \ \ \forall m    \\
& L_{m,s-1} + \Delta L \le L_{m,s} , \ \  \ \ \ \ \ \ \ \ \ \ \ \ \ \   \forall m,s \\
& \frac{s}{S}L_{m,\min} \le L_{m,s} \le \frac{s}{S}L_{m,\max},  \ \    \forall m  \\
& L_{m,s,n-1}+\Delta L_n \leq L_{m,s,n},   \ \ \ \ \ \ \ \ \ \   \! \! \! \forall m,s,n \label{eq:P3j} \\
&   L_{m,s-1} \le L_{m,s,n} \le L_{m,s}, \ \ \ \ \  \ \ \ \ \ \  \ \! \!\! \ \forall m,s,n  \label{eq:P3k}\\
&\mathcal{C}^0=0, \; \mathcal{C}^1=0, \ \ \ \ \ \ \  \ \ \ \  \ \ \ \ \ \ \ \ \  \  \ \ \ \ \ \ \  \forall m,s 
\\
&a_{m,s,n}\in\{0,1\}, \ \ \ \ \ \ \   \  \ \ \ \ \ \ \ \ \ \ \ \ \ \  \ \ \ \ \ \ \  \forall m,s,n 
\end{align}
\end{subequations}
where $\eta > 0$ denotes the cost coefficient for activating each intra-segment antenna, measured in bps/Hz per activated antenna. The constraint in \eqref{eq:P3j} is imposed to preserve the ordering of antennas after movement, i.e., antenna 1 cannot move to a position ahead of antenna 2. Constraint \eqref{eq:P3k} ensures that each antenna remains within its designated segment. In addition, the parameter set $\boldsymbol{\tilde \psi}$, which collects the parameters of all $M$ tentacles, is defined as $\boldsymbol{\tilde \psi}
\triangleq
\left\{\boldsymbol{\tilde \psi}_1, \boldsymbol{\tilde \psi}_2, \boldsymbol{\tilde \psi}_3\right\}_{m=1}^{M}
\in \mathbb{R}^{(3MS+M+2MS(N-1))}$. The motion parameter set $\boldsymbol{\tilde \psi}_1$ is given by \vspace{0.05cm}  $\boldsymbol{\tilde \psi}_1\!\triangleq\!
\left[
A_{m,1},\!..., A_{m,S},
v_{m,1},\!..., v_{m,S},
L_{m,1},\!..., L_{m,S},
\theta_{m}
\right]^{ T}\!$,  while the reconfigurable antenna position parameter set $\boldsymbol{\tilde \psi}_2$ is
\begin{equation}
\boldsymbol{\tilde \psi}_2
\triangleq
\left[ \ 
L_{m,1,1},..., L_{m,S,N-1}\ 
\right]^{ T}.
\end{equation}

The antenna activation set can be given by $\boldsymbol{\tilde \psi}_3
\triangleq
\left[ \ 
a_{m,1,1},..., a_{m,S,N-1}\ 
\right]^{ T},$
where \(a_{m,s,n} \in \{0,1\}\) indicates whether the \(n\)-th candidate intra-segment antenna on the \(s\)-th segment of the \(m\)-th tentacle is activated. Note that antennas located at segment endpoints require no activation variables, as they are always retained.

\section{Problem Solution of SEAC Scheme}

\label{scheme 2_solution}

Problem (P1) is non-convex due to multiple nonlinearities. Specifically, the objective contains a sum of logarithmic fractional terms that depend nonlinearly on the 3D antenna coordinates. Moreover, the antenna positions are determined by the segment motion-control parameters through an inherently nonlinear mapping. As shown in \eqref{3D location}, each antenna coordinate is a nonlinear function of parameters such as $v_{m,s}$ and $\theta_m$. In addition, \eqref{projection1} shows that the mapping from the projected segment length to the actual arc length is nonlinear.

\newcolumntype{Y}{>{\raggedright\arraybackslash}X}
\renewcommand\tabularxcolumn[1]{m{#1}}

\begin{table}[t]
\caption{Comparison of Optimization Methods Under the Two Schemes}
\label{tab:summary}
\centering
\footnotesize
\setlength{\tabcolsep}{2pt}
\renewcommand{\arraystretch}{1.15}
\begin{tabularx}{\columnwidth}{@{}
>{\centering\arraybackslash}m{0.13\columnwidth}
>{\raggedright\arraybackslash}m{0.31\columnwidth}
>{\raggedright\arraybackslash}X
>{\centering\arraybackslash}m{0.20\columnwidth}
@{}}
\toprule
\textbf{Scheme} &
\textbf{Coupling} &
\textbf{Proposed Method} &
\textbf{Convergence} \\
\midrule
SEAC &
Inter-segment continuity &
PDD-PGA &
Stationary point \\
\addlinespace[1mm]
HEIAC &
Segment shape, antenna positioning, and antenna activation &
BCD (PDD-PGA + PGA) + Greedy backward &
Stationary point \\
\bottomrule
\end{tabularx}
\end{table}

Beyond these nonlinearities, Problem (P1) also involves inter-segment coupling, as summarized in Table~\ref{tab:summary}. Specifically, the initial position and curvature of each segment depend on the terminal position and curvature of its preceding segment, requiring the continuity constraint in \eqref{eq:P2h} to ensure geometric smoothness, which further complicates Problem (P1).

To handle the nonlinear equality constraints, we adopt a penalty dual decomposition (PDD) framework. To begin, we present a geometric mapping model that reformulates the array-geometry optimization in terms of the motion-control parameters, as the antenna locations in the segment-wise SRA system are governed by segment motions. The mapping is given by
$\mathbf r=\mathbf g(\boldsymbol{\bar\psi})$, where
$\mathbf g:\mathbb R^{3MS+M}\rightarrow\mathbb R^{3MS}$.
Accordingly, the objective function can be equivalently expressed as
\begin{align}
R(\boldsymbol{\bar \psi}) = \sum_{k\in\mathcal{K}} \log_2(1+\gamma_{k}(\boldsymbol{\bar \psi})),
\end{align}
Next, we define the stacked  inter-segment continuity constraint vector as
\begin{equation}
\label{g_phi_bar}
\mathbf{p}(\boldsymbol{\bar \psi}) =
\begin{bmatrix}
\{ p_{m,s,0}(\boldsymbol{\bar \psi}) \}_{m=1,s=2}^{M,S} \\[1mm]
\{ p_{m,s,1}(\boldsymbol{\bar \psi}) \}_{m=1,s=2}^{M,S}
\end{bmatrix},
\end{equation}
where $ p_{m,s,0}(\boldsymbol{\bar \psi})
= \mathcal{C}^0$ and $p_{m,s,1}(\boldsymbol{\bar \psi})
= \mathcal{C}^1 $. By observing the expressions of $\mathcal{C}^0$ and $\mathcal{C}^1$ in \eqref{C_b}, we can conclude that the constraints are non-affine, involving multiplicative couplings among motion parameters and nonlinear functions. Such nonlinear equality constraints yield a nonconvex feasible set.

Consequently, we apply the PDD method in the outer loop, the inter-segment continuity and smoothness constraints $\mathcal{C}^0$ and $\mathcal{C}^1$ are incorporated into the penalty term of an augmented Lagrangian function. Specifically, the augmented Lagrangian function of Problem (P1) is given by
\begin{equation}
\label{Lagrangian}
\mathcal{L}(\boldsymbol{\bar \psi}, \boldsymbol{\lambda})
=
R(\boldsymbol{\bar \psi})
+
\boldsymbol{\lambda}^T \mathbf{p}(\boldsymbol{\bar \psi})
-
\frac{\rho}{2}
\|\mathbf{p}(\boldsymbol{\bar \psi})\|^2,
\end{equation}
where $\boldsymbol{\lambda}$ collects the dual variables associated
with the continuity constraints and $\rho>0$ denotes the penalty
parameter. Accordingly, Problem (P1) can be reformulated as 
\begin{subequations}\label{eq:P2}
\begin{align}
\text{(P1.1): } \nonumber \\ \!   
\max_{\boldsymbol{\bar \psi}}& \ \ \ \
 \mathcal{L}(\boldsymbol{\bar \psi}, \boldsymbol{\lambda})  \\
\ \ \ \ \text{s.t.}\ 
&\ \ \ \  0 \le A_{m,s} \le A_{\max}, \ \ \  \ \ \ \ \ \ \ \ \ \ \ \ \ \ \ \ \ \ \  \forall m,s \label{eq:P1.1b}\\
& \ \ \ \    0 \le v_{m,s} \le v_{\max}, \ \ \ \ \ \  \ \ \ \ \ \ \ \ \ \ \ \ \ \ \ \ \ \ \ \! \forall m,s \label{eq:P1.1c}\\
& \ \ \ \ \theta_{m-1}+\Delta \theta  \le \theta_m, \ \ \ \  \ \ \ \ \ \ \ \ \ \ \ \ \ \ \ \ \ \ \ \ \!    \forall m \label{eq:P1.1d}\\
& \ \ \ \  \theta_{m,\min}  \le \theta_m \le \theta_{m,\max}, \ \ \ \ \ \ \ \ \ \ \ \ \ \ \forall m  \label{eq:P1.1e}\\
& \ \ \ \ L_{m,s-1} + \Delta L \le L_{m,s} , \ \  \ \ \ \ \ \ \ \ \ \ \ \ \  \ \ \!  \forall m,s \label{eq:P1.1f}\\[-0.5mm]
& \ \ \ \ \frac{s}{S}L_{m,\min} \le L_{m,s} \le \frac{s}{S}L_{m,\max}, \  \    \forall m  \label{eq:P1.1g}
\end{align}
\end{subequations}
Since $\mathcal{L}(\boldsymbol{\bar \psi}, \boldsymbol{\lambda})$ is highly nonconcave and involves implicit geometric dependencies between the antenna coordinates and the motion-control parameters, we adopt the PGA method in the inner loop to approximately solve each PDD subproblem. Then, the gradient of the augmented objective is
\begin{equation}
    \nabla_{\boldsymbol{\bar \psi}}\mathcal{L}(\boldsymbol{\bar \psi}, \boldsymbol{\lambda})
=
\nabla_{\boldsymbol{\bar \psi}} R(\boldsymbol{\bar \psi})
+
\nabla_{\boldsymbol{\bar \psi}} \mathbf{p}(\boldsymbol{\bar \psi})^T(\boldsymbol{\lambda}
-
\rho\, \mathbf{p}(\boldsymbol{\bar \psi})),
\end{equation}
where $\nabla_{\boldsymbol{\bar \psi}} R(\boldsymbol{\bar \psi})$ and $\nabla_{\boldsymbol{\bar \psi}} \mathbf p(\boldsymbol{\bar \psi})$ denote the gradients of the rate and penalty terms with respect to $\boldsymbol{\bar \psi}$, respectively. These gradients can be derived via the chain rule by accounting for the nonlinear mapping between the antenna coordinates and the motion-control parameters, and are omitted for brevity.

Using the above gradient, the PGA method updates the motion-control parameters through iterative gradient ascent and projection steps. At the $\rm{i}$-th PGA iteration, given the current point  $\boldsymbol{\bar \psi}^{(\rm{i})}$, we first perform a gradient ascent step:

\begin{equation}
{\boldsymbol{ \bar \psi}}^{(\rm{i}+\frac{1}{2})}
=
\boldsymbol{\bar \psi}^{(\rm{i})}
+
\eta_{\rm{i}}
\nabla_{\boldsymbol{\bar \psi}}\mathcal{L}(\boldsymbol{\bar \psi}^{(\rm{i})}, \boldsymbol{\lambda}),
\end{equation}
where ${\boldsymbol{\bar \psi}}^{(\rm{i}+\frac{1}{2})}$ denotes the intermediate variable after the gradient ascent step, and $\eta_{\rm{i}} > 0$ is the step size. To ensure feasibility after the gradient ascent step, the intermediate point $\bar{\boldsymbol{\psi}}^{(i+\frac{1}{2})}$ is projected onto the feasible set as

\begin{equation}
\label{PGA_P2}
\boldsymbol{\bar{\psi}}^{(\rm{i}+1)}
=
\Pi_{\mathcal D}\!\left(
\bar{\boldsymbol{\psi}}^{(\rm{i}+\frac{1}{2})}
\right).
\end{equation}

\begin{algorithm}[!t]
\caption{BCD–PDD–PGA for Continuous Variables' Optimization Under a Fixed Activation Pattern}
\label{alg:BCD_P3}
\centering
\small
\begin{algorithmic}[1] 
\\
\textbf{Input:} Motions parameters $ \boldsymbol{\tilde \psi}_1^{(0)} = \{A_{m,s}^{(0)}, v_{m,s}^{(0)}, L_{m,s}^{(0)}, \theta_m^{(0)}\}$ $\forall m,s$ from SEAC solution; uniformly initialized per-antenna position parameters $\boldsymbol{\tilde \psi}_2^{(0)}=\{L_{m,s,n}^{(0)}\}$ $\forall m,s,n$; thresholds $\varepsilon_{\mathrm{out}}, \varepsilon_{\mathrm{in}} > 0$; $ \lambda^{(0)}=0, \rho^{(0)}=1$.\\
\textbf{Output:} $\{\hat A_{m,s}, \hat v_{m,s},\hat L_{m,s}, \hat \theta_m, \hat L_{m,s,n}\}$.

\Repeat

\State \textbf{Block 1:Fix activated antenna positions; optimize motion parameters.}

\State Fix $\boldsymbol{\tilde \psi}^{(\mathrm{z})}_2=\{L_{m,s,n}^{(\mathrm{z})}\}$ $\forall m,s,n$.

\State Run PDD-PGA inner loop:

\Repeat
\Repeat
\State Gradient step and projection:
\[
\boldsymbol{\tilde{\psi}}_1^{(\mathrm{i}+1)}
=
\Pi_{\mathcal D}\!\left(
\boldsymbol{\tilde{\psi}}^{(\mathrm{i})}_1
+
\eta_{\mathrm{i}}
\nabla_{\boldsymbol{\tilde{\psi}}_1}
\mathcal{L}(\boldsymbol{\tilde{\psi}}_1^{(\mathrm{i})},\boldsymbol{\lambda}, \boldsymbol{\tilde{\psi}}_2^{(\mathrm{z})})
\right)\]
\Until{$\|\boldsymbol{\tilde \psi}_1^{(\mathrm{i}+1)} - \boldsymbol{\tilde \psi}_1^{(\mathrm{i})}\| \leq \varepsilon_{\mathrm{in}}$}
\State Update dual variables:
\[
\boldsymbol{\lambda}^{(\rm{t}+1)}
=
\boldsymbol{\lambda}^{(\rm{t})}
-
\rho\,\mathbf{p}(\boldsymbol{\tilde  \psi}_1^{(\rm{t}+1)}).
\]
\Until{$\|\mathbf{p}(\boldsymbol{\tilde  \psi}_1^{(\rm{t}+1)})\| \le \epsilon_{\mathrm{pdd}}$}

\vspace{0.1cm}

\State Set $\{A_{m,s}^{(\mathrm{z}+1)}, v_{m,s}^{(\mathrm{z}+1)}, \theta_m^{(\mathrm{z}+1)}, L_{m,s}^{(\mathrm{z}+1)}\}
\leftarrow \boldsymbol{\tilde \psi}_1^{(\mathrm{converged})}$.

\vspace{1mm}
\State \textbf{Block 2: Fix motion parameters; optimize activated antenna positions.}

\State Fix $\boldsymbol{\tilde \psi}^{(\mathrm{z})}_1=\{A_{m,s}^{(\mathrm{z})}, v_{m,s}^{(\mathrm{z})},  L_{m,s}^{(\mathrm{z})},\theta_m^{(\mathrm{z})}\}$ $\forall m,s$.

\State Run PGA inner loop:

\Repeat
\State Gradient step and projection:
\[
\boldsymbol{\tilde{\psi}}_2^{(\mathrm{i}+1)}
=
\Pi_{\mathcal D}\!\left(
\boldsymbol{\tilde{\psi}}^{(\mathrm{i})}_2
+
\eta_{\mathrm{i}}
\nabla_{\boldsymbol{\tilde{\psi}}_2}
R(\boldsymbol{\tilde{\psi}}_1^{(\mathrm{z})},\boldsymbol{\tilde{\psi}}_2^{(\mathrm{i})})
\right)\]
\Until{$\|\boldsymbol{\tilde \psi}_2^{(\mathrm{i}+1)} - \boldsymbol{\tilde \psi}_2^{(\mathrm{i})}\| \leq \varepsilon_{\mathrm{in}}$}
\vspace{1mm}
\State Set $\{L_{m,s,n}^{(\mathrm{z}+1)}\}
\leftarrow \boldsymbol{\tilde \psi}_2^{(\mathrm{converged})}$.
\vspace{1mm}
\Until{$\big| \boldsymbol{\tilde \psi}^{(\mathrm{z}+1)} - \boldsymbol{\tilde \psi}^{(\mathrm{z})} \big| \leq \varepsilon_{\mathrm{out}}$}

\end{algorithmic}
\end{algorithm}

In particular, the feasible set $\mathcal D$ can be divided into ordering constraints and box constraints. The ordering constraints \eqref{eq:P1.1d} and \eqref{eq:P1.1f} are enforced via sequential forward-backward projection steps, whereas the projection onto the box constraints \eqref{eq:P1.1b}, \eqref{eq:P1.1c}, \eqref{eq:P1.1e} and \eqref{eq:P1.1g} admits a closed-form expression for each $m,s$, the update of $\bar{\psi}_{m,s}$ is obtained via
\begin{equation}
    \bar{\psi}_{m,s}^{(\rm{i}+1)} = \min\!\left\{\max\!\left\{\bar{\psi}_{m,s}^{(\rm{i}+\frac{1}{2})},\ \bar{\psi}_{m,s,\min}\right\},\ \bar{\psi}_{m,s,\max}\right\}.
\end{equation}
where $\bar{\psi}_{m,s,\min}$ and $\bar{\psi}_{m,s,\max}$ represent the upper and lower bounds, respectively. 

The inner PGA loop is terminated when the variation between two consecutive primal iterates becomes sufficiently small, i.e.,

\begin{equation}
\left\|\bar{\boldsymbol{\psi}}^{(\rm{i}+1)}-\bar{\boldsymbol{\psi}}^{(\rm{i})}\right\| < \epsilon_{\rm{in}},
\end{equation}
where $\epsilon_{\rm{in}}$ denotes the inner loop threshold. After the inner PGA loop converges at the $\rm{t}$-th PDD iteration, the obtained solution is denoted by $\bar{\boldsymbol{\psi}}^{(\rm{t}+1)}$, and the dual variable is updated as

\begin{equation}
\boldsymbol{\lambda}^{(\rm{t}+1)}
=
\boldsymbol{\lambda}^{(\rm{t})}
-
\rho \mathbf{p}\big(\bar{\boldsymbol{\psi}}^{(\rm{t}+1)}\big).
\end{equation}

By alternating between the primal and dual updates, the PDD framework promotes constraint satisfaction while improving the objective value. The outer loop terminates when the constraint residual satisfies $\|\mathbf{p}\big(\bar{\boldsymbol{\psi}}^{(\rm{t}+1)}\big)\|
\leq
\epsilon_{\rm pdd}$. Otherwise, the penalty parameter $\rho$ is adaptively updated, with an upper bound $\rho_{\max}$, to balance objective improvement and constraint enforcement.

\section{Problem Solution of HEIAC Scheme}
Compared with problem (P1), problem (P2) further optimizes the intra-segment antenna positions and the number of antennas per segment. Based on the nature of the optimization variables, we partition them into two categories: binary variables $a_{m,s,n}$ and continuous variables, such as $A_{m,s}$ and $L_{m,s,n}$. Due to the nonlinear coupling among antenna activation, intra-segment antenna repositioning, and segment-shape control, as summarized in Table \ref{tab:summary}, problem (P2) is a mixed-integer non-convex problem and is difficult to solve directly. To maintain a concise and tractable solution framework, the optimization are decoupled into two parts. \textbf{Part I} optimizes the continuous variables with fixed binary variables, which specify the antenna activation pattern, while \textbf{Part II} optimizes the binary variables.

\subsection{Motion and Antenna Repositioning Variable Optimization}
\label{Scheme 3 Part I}
In Part I, compared to Problem (P1), we introduce the parameter $L_{m,s,n}$ to change the MA antenna position individually, which substantially increases the optimization dimension, especially in dense-antenna scenarios. Moreover, since the antennas move along the arc length of each segment, the antenna repositioning within a segment is strongly coupled with the segment shape determined by the motion parameters.

To solve this challenged problem, we adopt a block coordinate descent (BCD) strategy within the proposed PDD-PGA framework. By using the BCD decomposition, we partition the variables into \textbf{block 1}: the motion-parameter block and \textbf{block 2}: the reconfigurable antenna position parameter block, and update the blocks sequentially at the inner iterations. 

To accelerate convergence, we adopt the SEAC solution as the initial point for the motion parameters. In addition, the intra-segment antenna positions are initialized with uniform spacing along each segment, while satisfying the constraints in \eqref{eq:P3j} and \eqref{eq:P3k}. The proposed algorithm adopts a two-layer iterative structure with one outer BCD loop and two inner loops. Specifically, each block is updated through an inner iteration indexed by $\mathrm{i}$, while the outer iteration is indexed by $\mathrm{z}$. Specially, in the motion-parameter block, the PDD loop is indexed by $\mathrm{t}$. The overall procedure is summarized in Algorithm~1. At the $(\mathrm{z}\!+\!1)$-th outer iteration, we first optimize the motion-parameter block $\boldsymbol{\tilde \psi}_1 = \{ \{\!A_{m,s}\}_{m=1,s=2}^{M,S}, \! \{v_{m,s}\}_{m=1,s=2}^{M,S},\! \{L_{m,s}\}_{m=1,s=2}^{M,S},\! \{\theta_{m}\}_{m=1}^{M} \}$ as
\begin{subequations}\label{eq:P2.2}
\begin{align}
\text{(P2.1): } 
\max_{\boldsymbol{\bar \psi}_1}& \ \ \ \
\mathcal{L}(\boldsymbol{\tilde \psi}_1, \boldsymbol{\lambda}, \boldsymbol{\tilde \psi}^{(\mathrm{z})}_2) \label{eq:P3.1a}  \\
\ \ \ \ \text{s.t.}\ 
&\ \ \ \  0 \le A_{m,s} \le A_{\max}, \ \ \ \ \ \ \ \  \ \ \ \!  \forall m,s \label{eq:P3.1b}\\
& \ \ \ \ 0 \le v_{m,s} \le v_{\max}, \ \ \ \ \ \ \ \ \ \  \ \   \forall m,s \label{eq:P3.1c} \\
&\ \ \ \  \theta_{m-1}+\Delta \theta  \le \theta_m, \  \  \ \  \ \ \ \  \ \ \ \forall m \label{eq:P3.1d} \\
&\ \ \ \ \theta_{m,\min}  \le \theta_m \le \theta_{m,\max},  \ \  \forall m \label{eq:P3.1e} \\
& \ \ \ \ L_{m,s-1} + \Delta L \le L_{m,s} ,  \ \ \ \   \forall m,s \\[-0.5mm]
& \ \ \ \ \frac{s}{S}L_{m,\min} \le L_{m,s} \le \frac{s}{S}L_{m,\max}, \ \! \! \    \forall m
\end{align}
\end{subequations}
where the intra-segment antenna position variable block \vspace{0.08cm}$\boldsymbol{\tilde \psi}_2 = { \{  L_{m,s,n}\} }_{m=1,s=1,n=1}^{M,S,N-1}$ has been obtained in the $\mathrm{z}$-th BCD iteration and is treated as fixed. The term $-\eta \sum a_{m,s,n}$ is constant with respect to the continuous variables under a fixed activation pattern, and is therefore dropped from Problem (P2.1) without affecting its solution.
%It is worth noting that the term $-\eta \sum a_{m,s,n}$ is constant with respect to the continuous variables under a fixed activation pattern, and is therefore dropped from the inner BCD--PDD--PGA procedure without affecting its solution.
Since $\mathcal{L}(\boldsymbol{\tilde \psi}_1, \boldsymbol{\lambda}, \boldsymbol{\tilde \psi}_2^{(\mathrm{z})})$ is non-concave, we also employ the PGA method, which consists of a gradient ascent step followed by a projection onto the feasible set as
\begin{equation}
\label{PGA_P3}
\boldsymbol{\tilde{\psi}}_1^{(\mathrm{i}+1)}
=
\Pi_{\mathcal D}\!\left(
\boldsymbol{\tilde{\psi}}^{(\mathrm{i})}_1
+
\eta_{\mathrm{i}}
\nabla_{\boldsymbol{\tilde{\psi}}_1}
\mathcal{L}(\boldsymbol{\tilde{\psi}}_1^{(\mathrm{i})},\boldsymbol{\lambda},\boldsymbol{\tilde{\psi}}_2^{(\mathrm{z})})
\right),
\end{equation}
where the projection method is similar to that of Section \ref{scheme 2_solution}. After the PGA procedure converges, we obtain a solution denoted by ${\boldsymbol{\tilde  \psi}_1}^{(\rm{t}+1)}$, and then dual variables are updated as 
\begin{equation}
\boldsymbol{\lambda}^{(\rm{t}+1)}
=
\boldsymbol{\lambda}^{(\rm{t})}
-
\rho\,\mathbf{p}(\boldsymbol{\tilde  \psi}_1^{(\rm{t}+1)}).
\end{equation}

Afterwards, we update per-antenna position variable block $\boldsymbol{\tilde \psi}_2$ while fixing the motion-parameter block $\boldsymbol{\tilde \psi}^{(\mathrm{z})}_1$
at the $(\mathrm{z}\!+\!1)$-th outer iteration. Since this subproblem does not involve the nonlinear segment-continuity constraints, it can be directly solved by PGA as follows::
\begin{subequations}
\begin{align}
\text{(P2.2): }  
 \max_{\boldsymbol{\tilde \psi}_2} & \   \ \ \ \  
{R}(\boldsymbol{\tilde \psi}_2, \boldsymbol{\tilde \psi}^{(\mathrm{z})}_1)
\\
 \text{s.t.} 
&\ \ \ \ \ \  L_{m,s,n-1}+\Delta L_n \leq L_{m,s,n}, \ \ \! \! \forall m,s,n \label{eq:P2.2b}\\
& \ \ \ \  \ \ L_{m,s-1} \le L_{m,s,n} \le L_{m,s},  \ \ \  \forall m,s,n \label{eq:P2.2c}  
\end{align}
\end{subequations}
where the motion-parameter block has been obtained at the $\mathrm{z}$-th outer iteration. To solve problem (P2.2), we perform a gradient ascent step followed by a projection onto the feasible set as
\begin{equation}
\label{PGA_P4}
\boldsymbol{\tilde{\psi}}_2^{(\mathrm{i}+1)}
=
\Pi_{\mathcal D}\!\left(
\boldsymbol{\tilde{\psi}}^{(\mathrm{i})}_2
+
\eta_{\mathrm{i}}
\nabla_{\boldsymbol{\tilde{\psi}}_2}
R(\boldsymbol{\tilde{\psi}}_2^{(\mathrm{i})},\boldsymbol{\tilde{\psi}}_1^{(\mathrm{z})})
\right),
\end{equation}
where the ordering constraint \eqref{eq:P2.2b} is also enforced using a sequential forward-backward projection step, whereas the projection onto the box constraint \eqref{eq:P2.2c} is implemented as follows
\begin{align}
\label{projection4}
L_{m,s,n}^{(\mathrm{i}+1)}
=
\min\!\Big\{
\max\big\{{L}_{m,s,n}^{(\mathrm{i}+\frac{1}{2})},\, L_{m,s-1}\big\},
\, L_{m,s}
\Big\},
\end{align}
where ${L}_{m,s,n}^{(\mathrm{i}+\frac{1}{2})}$ denotes the intermediate point. The projection in \eqref{projection4} follows directly from the fact that $L_{m,s-1}$ and $L_{m,s}$ are updated in the motion-parameter block and are therefore treated as known constants in the current step.

\begin{algorithm}[!t]
\centering
\begin{algorithmic}[1] 
\\
\textbf{Input:} Greedy iteration index $\tau=0$; inital activation states $a_{m,s,n}^{(0)}=1$, $\forall m,s,n$\\
\textbf{Run} Algorithm 1 under $\mathbf{a}^{(\tau)}$ to obtain the corresponding solution for continuous variables and the utility $\Phi(\mathbf{a}^{(\tau)})$;
\\
\textbf{Output:} Optimized $\{A_{m,s}^{\star}, v_{m,s}^{\star}, \theta_m^{\star}, L_{m,s}^{\star}, L_{m,s,n}^{\star}, \mathbf{a}^{\star}\}$.
  \Repeat
    \For{each active intra-segment antenna $(m,s,n)$ with $a_{m,s,n}^{(\tau)}=1$}
      \State Construct $\mathbf{a}_{-(m,s,n)}^{(\tau)}$ by deactivating antenna $(m,s,n)$;
      \State Run Algorithm 1 under $\mathbf{a}_{-(m,s,n)}^{(\tau)}$;
      \State Compute
      $G_{m,s,n}^{(\tau)}=\Phi\!\left(\mathbf{a}_{-(m,s,n)}^{(\tau)}\right)-\Phi\!\left(\mathbf{a}^{(\tau)}\right)$;
    \EndFor
    \State Select
    $(m^\star,s^\star,n^\star)=\arg\max_{a_{m,s,n}^{(\tau)}=1} G_{m,s,n}^{(\tau)}$;
    \If{$G_{m^\star,s^\star,n^\star}^{(\tau)} \ge 0$}
      \State Update $\mathbf{a}^{(\tau+1)}=\mathbf{a}_{-(m^\star,s^\star,n^\star)}^{(\tau)}$;
      \State Run Algorithm 1 under $\mathbf{a}^{(\tau+1)}$;
      \State Update $\Phi(\mathbf{a}^{(\tau+1)})$ and $\tau=\tau+1$;
    \Else
      \State Terminate;
    \EndIf
  \Until
  \State No active antenna yields a nonnegative removal gain.
\end{algorithmic}
\caption{Greedy Backward Antenna Selection for Problem (P2)}
\end{algorithm}

We alternately optimize the aforementioned two blocks until the algorithm satisfies the following convergence criterion
\begin{equation}
\big| \boldsymbol{\tilde \psi}^{(\mathrm{z}+1)} - \boldsymbol{\tilde \psi}^{(\mathrm{z})} \big| \leq \varepsilon_{\mathrm{out}},
\end{equation}
where $\varepsilon_{\mathrm{out}}$ indicates the threshold.

\subsection{Binary Variables Optimization}
\label{Scheme 3 Part II}
In Part II, building upon the optimized segment shapes and antenna positions obtained in Part I, we further employ a greedy backward antenna selection strategy to determine the binary antenna activation variables $a_{m,s,n} \ \forall \ \{m,s,n\}$, thereby jointly optimizing the array configuration and the number of activated antennas. Starting from full activation, we iteratively deactivate the intra-segment antenna whose removal yields the largest utility gain until no further improvement is achieved.

At the $\tau$-th greedy iteration, \vspace{0.05cm} the antenna activation pattern is 
$\mathbf{a}^{(\tau)} \triangleq [a_{m,s,n}^{(\tau)}]_{m,s,n}$, where the initialization $a_{m,s,n}^{(0)}=1$, $\forall(m,s,n)$, corresponds to the fully activated configuration. We further define the candidate activation pattern $\mathbf{a}_{-(m,s,n)}^{(\tau)}$, which is obtained from $\mathbf{a}^{(\tau)}$ by setting the entry associated with antenna $(m,s,n)$ to zero while keeping all other entries unchanged. The removal gain of antenna $(m,s,n)$ is defined as
\begin{equation}
G_{m,s,n}^{(\tau)} \triangleq \Phi\!\left(\mathbf{a}_{-(m,s,n)}^{(\tau)}\right) - \Phi\!\left(\mathbf{a}^{(\tau)}\right),
\end{equation}
where the utility function $\Phi(\cdot)$ is given by
\begin{equation}
\label{utility function}
\Phi\!\left(\mathbf{a}^{(\tau)}\right) \triangleq \hat R\!\left(\mathbf{a}^{(\tau)}\right) 
- \eta \sum_{m=1}^{M}\sum_{s=1}^{S}\sum_{n=1}^{N-1} a_{m,s,n}^{(\tau)},
\end{equation}
in which $ \hat R(\mathbf{a}^{(\tau)})$ denotes achieved sum-rate returned by the proposed inner BCD--PDD--PGA solver in Section~\ref{Scheme 3 Part I} under 
the fixed activation pattern $\mathbf{a}^{(\tau)}$. Accordingly, $G_{m,s,n}^{(\tau)} > 0$ 
indicates that deactivating antenna $(m,s,n)$ improves the utility, whereas 
$G_{m,s,n}^{(\tau)} < 0$ implies that its removal is detrimental.

\begin{table}[t]
\centering
\caption{Penalty residual comparison of PGA and SCA.}
\label{tab:pdd_penalty_compare}
\renewcommand{\arraystretch}{1.15}
\setlength{\tabcolsep}{10pt}
\begin{tabular}{|c|c|c|}
\hline
\textbf{SNR (dB)} & \textbf{PGA} & \textbf{SCA} \\
\hline
0  & $5.7633\times10^{-3}$ & $3.0400\times10^{-1}$ \\
3  & $6.2000\times10^{-3}$ & $3.8675\times10^{-1}$ \\
6  & $3.8154\times10^{-3}$ & $5.5075\times10^{-1}$ \\
9  & $4.5061\times10^{-3}$ & $6.0200\times10^{-1}$ \\
12 & $7.8271\times10^{-3}$ & $2.9378\times10^{-1}$ \\
15 & $4.0211\times10^{-3}$ & $1.9653\times10^{-1}$ \\
18 & $1.2697\times10^{-4}$ & $4.0477\times10^{-1}$ \\
21 & $1.6783\times10^{-3}$ & $3.1747\times10^{-1}$ \\
\hline
\end{tabular}
\vspace{-0.1cm}
\end{table}

Among all currently active intra-segment antennas, the best removal candidate is selected as
\begin{equation}
(m^\star,s^\star,n^\star)
=
\arg\max_{a_{m,s,n}^{(\tau)}=1}
G_{m,s,n}^{(\tau)},
\end{equation}
where at the $\tau$-th iteration, $G_{m,s,n}^{(\tau)}$ is evaluated for each active 
antenna $(m,s,n)$ with $a_{m,s,n}^{(\tau)}=1$, and $(m^\star,s^\star,n^\star)$ 
identifies the one whose removal yields the greatest utility gain. %In addition, a one-at-a-time deletion strategy is adopted to limit the computational complexity, while properly accounting for the coupled effect of each removal on the intra-segment antenna correlation and the continuous optimization subproblem in Section~\ref{Scheme 3 Part I}.
If the maximum removal gain is nonnegative, i.e., $G_{m^\star,s^\star,n^\star}^{(\tau)} \geq 0$, the removal is accepted, and the activation pattern is updated as
\begin{equation}
\mathbf{a}^{(\tau+1)}
=
\mathbf{a}_{-(m^\star,s^\star,n^\star)}^{(\tau)}.
\end{equation}
Otherwise, the greedy backward procedure terminates. 

The backward strategy is adopted because the optimal number of active antennas is unknown a priori. Starting from full activation avoids prematurely discarding useful antennas, while pruning removes antennas whose marginal gains are outweighed by activation cost and correlation. To reduce complexity, each greedy iteration uses a screening-and-refinement strategy: a lightweight pass evaluates all active candidates with a limited iteration budget, and the full inner solver is applied to the best candidate, reducing cost while mitigating incorrect pruning.

\section{Numerical Results}
\label{Numerical Results}
In this section, we evaluate the achievable sum-rate performance of the segment-wise SRA with the proposed SEAC and HEIAC schemes under various scenarios. Unless otherwise specified, the system parameters are $K=7$, $A_{\max}=0.2\lambda$, and $v_{\max}=5\lambda^{-1}$. The SEAC scheme adopts an SRA configuration with $M=4$ and $S=3$, whereas the HEIAC scheme uses $M=2$, $S=2$, and $N=2$. Thus, each HEIAC segment contains one intra-segment antenna and one segment-end antenna. The initial inter-antenna spacing along each tentacle is set to $0.1\lambda$, consistent with \cite{Wong2024}. The length of the $m$-th tentacle is constrained within $[L_m,4L_m]$, i.e., $L_{m,\min}=L_m$ and $L_{m,\max}=4L_m$. Its angular sweeping range is $\theta_{m,\min}=2\pi(m-1)/M$ and $\theta_{m,\max}=2\pi m/M$. Furthermore, the carrier frequency is $f_c=1.3$ GHz, and all results are averaged over 1000 independent channel realizations. In the considered settings, the outer BCD loop converges within 15 iterations on average, the PDD loop requires about 12 iterations, and each PGA loop converges within 100 iterations on average.

\subsection{Performance Comparison With Benchmark Schemes}
\label{Benchmark comparison}
To comprehensively evaluate the proposed schemes, we consider three concentric circular antenna array (CCAA)-based benchmarks: 1) a fixed CCAA \cite{Sun2018}, corresponding to the undeformed soft robot, as shown in Fig. \ref{Projection}; 2) a 2D reconfigurable CCAA (2D R-CCAA), where antennas move along circular tracks; and 3) a 3D reconfigurable CCAA (3D R-CCAA), where antennas move along circular tracks and the tracks are translated along the $z$-axis. The 2D R-CCAA extends the movable uniform circular array (UCA) in \cite{Basbug2017} to multi-ring CCAA, while the 3D R-CCAA further extends the movable cylindrical array in \cite{Guo2025} by enabling vertical movement of multiple rings with different radii. Note that in the undeformed state, the SRA and all benchmarks are initialized with identical antenna positions and array geometry, thereby ensuring that fair comparison.

%{
%\begin{figure*}[!t]
%\centering
%\subfloat[SEAC scheme.]{
    %\includegraphics[width=0.48\textwidth]{figs/Fig1_V1.eps}
    %\label{fig:seac}
%}
%\hfill
%\subfloat[HEIAC scheme.]{
    %\includegraphics[width=0.48\textwidth]{figs/Fig2_V1.eps}
    %\label{fig:heiac}
%}
%\caption{Sum-rate comparison of SRA, 3D R-CCAA, 2D R-CCAA, and CCAA under different SNR levels for SEAC scheme.}
%\label{fig:sumrate_compare}
%\end{figure*}
%}

 \begin{figure}[ptb]
\centering
{\includegraphics  [height=2.53in, width=3.4in]{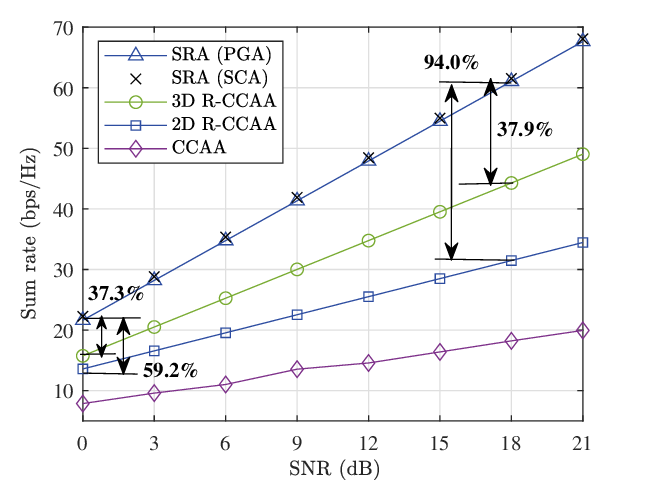}}
\caption{Sum-rate comparison of SRA, 3D R-CCAA, 2D R-CCAA, and CCAA under different SNR levels for SEAC scheme.}
\color{black}
\label{fig:sumrate_compare1}
\end{figure}

 \begin{figure}[ptb]
\centering
{\includegraphics  [height=2.53in, width=3.4in]{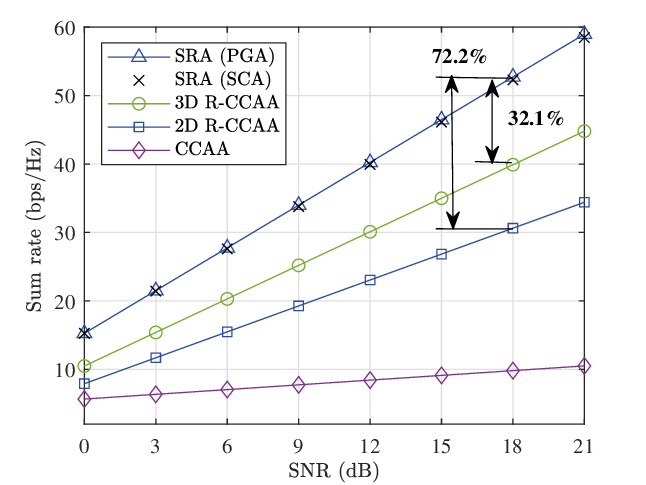}}
\caption{Sum-rate comparison of SRA, 3D R-CCAA, 2D R-CCAA, and CCAA under different SNR levels for HEIAC scheme.}
\color{black}
\label{fig:sumrate_compare2}
\end{figure}

In Fig.~\ref{fig:sumrate_compare1}, the sum-rate performance of the proposed SRA architecture under the SEAC scheme is evaluated against the 3D R-CCAA, 2D R-CCAA, and fixed CCAA benchmarks. Although PGA is used for SRA sum-rate optimization, the performance of successive convex approximation (SCA) is also included for comparison.  Fig.~\ref{fig:sumrate_compare1} indicates that although SCA achieves a slightly higher converged sum rate than PGA, it is significantly inferior in reducing the penalty residual \( \|\mathbf{p}\| \). As shown in Table \ref{tab:pdd_penalty_compare}, PGA achieves a penalty residual about two orders of magnitude lower than SCA, since it updates the variables based on the true augmented objective, whereas SCA relies on a local first-order surrogate that may not accurately capture the nonlinear constraint violation in our highly coupled problem. Compared with the SRA, the fixed CCAA achieves the worst performance, indicating that its static geometry limits spatial-correlation mitigation, thereby reducing the performance gains enabled by antenna repositioning. The 2D R-CCAA improves the sum rate by enabling antenna movement along circular tracks, while the 3D R-CCAA further enhances performance through additional $z$-axis translation for more flexible 3D adjustment. The proposed SRA achieves the highest sum rate. For instance, at an $\mathrm{SNR}=18$~dB, it outperforms the 3D R-CCAA and 2D R-CCAA by $37.9\%$ and $94.0\%$, respectively. This gain stems from its flexible array-geometry reconfiguration and effective-aperture enlargement enabled by segment-wise deformation and elongation, which jointly facilitate spatial-correlation mitigation. 

 \begin{figure}[ptb]
\centering
{\includegraphics  [height=2.53in, width=3.4in]{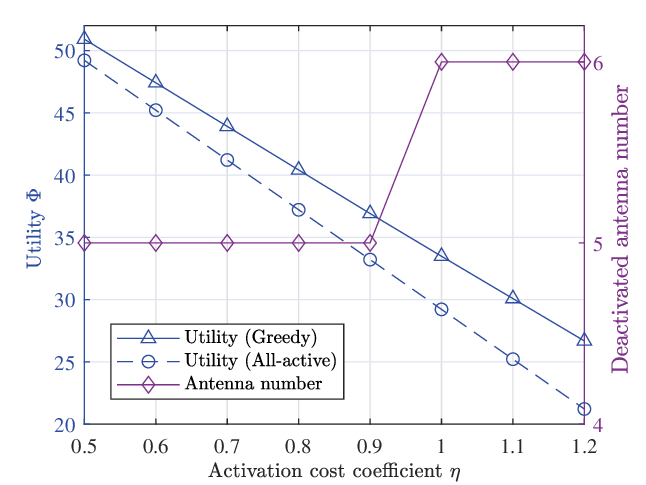}\vspace{-0.1cm}}
\caption{Performance of antenna activation in the HEIAC scheme (the left and right y-axes denote utility and the number of deactivated antennas).}
\color{black}
\label{antenna activation}
\end{figure}

 \begin{figure}[ptb]
\centering
{\includegraphics  [height=2.53in, width=3.4in]{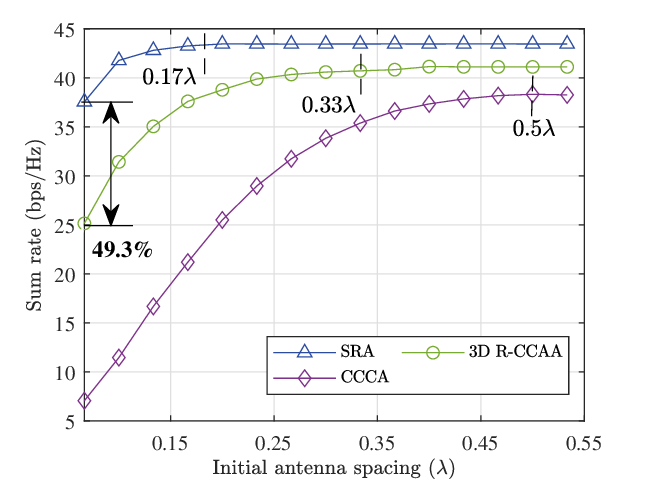}}
\caption{Performance comparison of SRA, 3D R-CCAA, and CCAA with different initial antenna spacings at an SNR of 9 dB.}
\color{black}
\label{array aperture}
\end{figure}

In Fig.~\ref{fig:sumrate_compare2}, we compare the sum-rate performance of the proposed SRA with those of the 3D R-CCCA, 2D-R-CCCA, and fixed CCAA under the HEIAC scheme, where movable intra-segment antennas are introduced in addition to the segment-end antennas. By jointly optimizing the continuous deformation of the soft robotic tentacles and the intra-segment antenna positions, the proposed SRA with the HEIAC scheme consistently achieves superior performance over all benchmark schemes. For example, it outperforms the 3D R-CCCA and 2D-R-CCCA by $32.1\%$ and $72.2\%$, respectively.

Fig.~\ref{antenna activation} evaluates the greedy backward antenna activation in Algorithm~2 at $\mathrm{SNR}=15$ dB with $N=11$. Several antennas are deactivated among the \(MN=44\) candidates, indicating redundancy in the intermediate intra-segment antenna deployment. The optimal number of active antennas depends on the cost coefficient $\eta$: five antennas are removed for $\eta\in[0.5,0.9]$ and six for $\eta\in[1.0,1.2]$, indicating that a larger $\eta$ yields sparser activation. Compared with the all-active baseline, the greedy scheme achieves a higher utility value by deactivating ineffective antennas, where the utility is defined in \eqref{utility function} as the difference between the sum rate and the antenna activation cost. In addition, although the sum-rate curves are omitted for clarity in Fig.~\ref{antenna activation}, the rate loss is marginal. For example, at $\eta=1.2$, the greedy scheme deactivates six antennas, reducing the sum rate from $69.2091$ to $67.4763$ bps/Hz, i.e., by only about $2.5\%$, while eliminating $15\%$ of the candidate intra-segment antennas. This highlights the effectiveness of the greedy backward antenna activation method for the HEIAC scheme.

Fig.~\ref{array aperture} compares the SRA, 3D R-CCAA, and CCAA with different initial inter-element spacings. The SRA with SEAC is considered as a representative case, as the HEIAC-based SRA exhibits a similar trend. The spacing is varied by adjusting the array radius; for the SRA, this is achieved by changing the initial tentacle length. As shown in Fig.~\ref{array aperture}, when the CCAA spacing exceeds about $0.5\lambda$ (i.e., radius larger than $1.5\lambda$ for $S=3$), spatial correlation becomes negligible, which agrees with classical theory. For the 3D R-CCAA, negligible correlation is achieved when the inter-element spacing exceeds $0.33\lambda$, indicating that antenna mobility along both circular tracks and the $z$-axis improves correlation mitigation. In contrast, the SRA suppresses spatial correlation even at an initial spacing of $0.17\lambda$, demonstrating superior performance for compact arrays. This gain arises from both stretching-enabled aperture enlargement and array-geometry reconfiguration enabled by
various motions.

\subsection{Performance Gain Analysis}
After comparing the proposed segment-wise SRA schemes with literature-based benchmarks, we consider the following SRA variants to reveal the sources of the gains: 4) SRA with extension--retraction motions only; 5) SRA with bending and sweeping motions only; and 6) SRA with non-independent control (NIC) for antenna repositioning, i.e., without segment-wise control or intra-segment reconfigurable antenna system.

\begin{figure}[ptb]
\centering
{\includegraphics  [height=2.53in, width=3.4in]{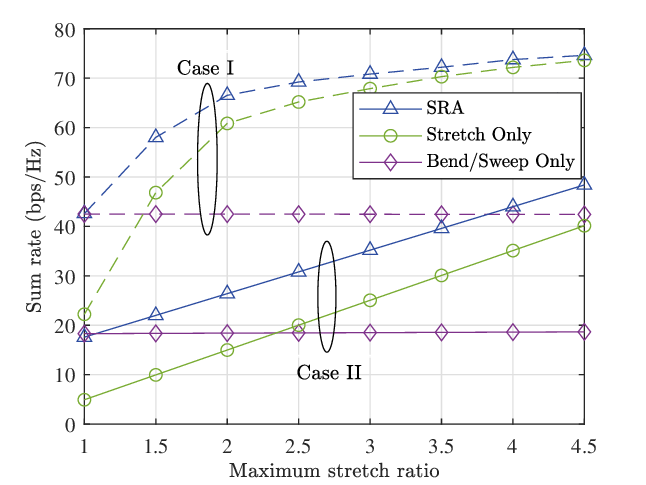}}
\caption{Contribution of different SRA motions to the sum-rate.}
\color{black}
\label{different SRA motions}
\end{figure}

 \begin{figure}[ptb]
\centering
{\includegraphics  [height=2.53in, width=3.4in]{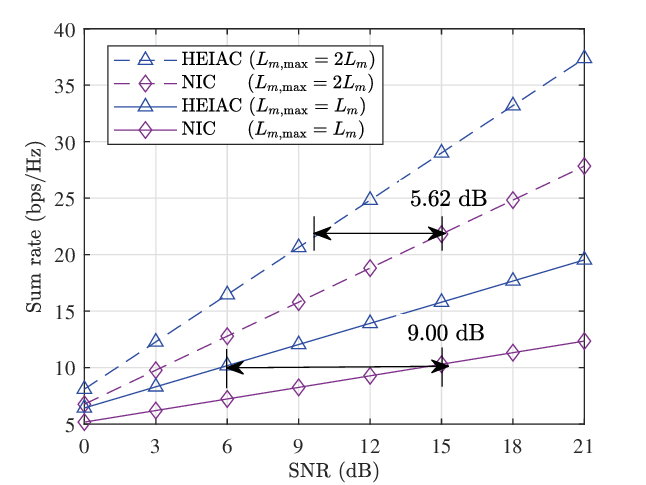}}
\caption{Comparison between the HEIAC scheme and the NIC benchmark.}
\color{black}
\label{NIC2}
\end{figure}

In Fig. \ref{different SRA motions}, the contributions of different SRA motions are evaluated by comparing the full-motion SRA with the SRA in extension–-retraction-only mode and the SRA in bending–-sweeping-only mode at $\mathrm{SNR}=15\ \mathrm{dB}$. The HEIAC-based SRA is used as a representative case, since the SEAC-based SRA shows a similar trend. Two cases are considered under the same aperture with an initial array radius of $0.4\lambda$: Case I, a dense antenna with $(M=4,S=3,N=3)$; and Case II, a sparse antenna with $(M=2,S=2,N=2)$. The result shows that, in both cases, when the maximum stretch ratio $L_{m,\max}$ is small, the full-motion SRA performs similarly to the bending--sweeping-only benchmark, since extension--retraction is limited, and the gain mainly comes from shape adaptation. As $L_{m,\max}$ increases, radial aperture expansion becomes dominant, while the role of extension--retraction differs between the two cases. In the dense antenna case, the stretch-only benchmark approaches the full-motion SRA at large $L_{m,\max}$, indicating that most gains come from extension--retraction, with bending and sweeping providing only marginal improvement. In contrast, in the sparse antenna case, the full-motion SRA still outperforms the stretch-only benchmark, indicating that bending and sweeping remain important through their synergy with stretching. This is because dense deployments are mainly constrained by small inter-antenna spacing and strong spatial correlation, where stretching effectively enlarges the effective aperture. In contrast, in sparse deployments, stretching alone provides limited gains in correlation mitigation, making bending and sweeping non-negligible for further array-geometry reconfiguration.

% \begin{figure}[ptb]
%\centering
%{\includegraphics  [height=2.53in, width=3.5in]{figs/Fig6.eps}}
%\caption{Performance comparison between the SEAC scheme and the NSC benchmark}
%\color{black}
%\label{NIC1}
%\end{figure}

In Fig.~\ref{NIC2}, the proposed SRA architecture is compared with the NIC benchmark using the HEIAC scheme as a representative example, since SEAC exhibits a similar trend. The results demonstrate that HEIAC significantly outperforms NIC, with more pronounced gains in the high-SNR regime. This is because the performance at low SNR is mainly noise-limited, whereas antenna correlation becomes more critical at high SNR, making array reconfiguration more beneficial. The gain over NIC is attributed to the additional DoFs provided by segment-wise control and the intra-segment MA mechanisms. Furthermore, comparing the one-fold and two-fold extension ranges indicates that NIC requires a higher SNR to achieve the same sum rate as HEIAC. For instance, to achieve sum rates of $10$ and $21.82$ bps/Hz, NIC requires $9$ and $5.62$ dB higher SNR than HEIAC in the one-fold and two-fold extension cases, respectively.

\subsection{Performance Evaluation Under Mutual Coupling}
Segment-wise SRAs can alleviate mutual coupling via various motions, similarly to mitigating spatial correlation. However, residual coupling in compact arrays may remain non-negligible and requires evaluation. Although advanced decoupling techniques can suppress inter-element coupling below $-20$ dB for $0.12\lambda$ spacing ~\cite{Wong2024}, we consider a compact SRA without such techniques and quantify mutual coupling using the model in \cite{Wong2024} for post-optimization evaluation, where the mutual coupling matrix $\boldsymbol{\Gamma}_{\rm mc}$ is given by $\boldsymbol{\Gamma}_{\rm mc}=\mathbf{Z}_{T}(\mathbf{Z}+\mathbf{Z}_{T})^{-1}$, where $\mathbf{Z}_{T}$ is the termination impedance matrix, and $\mathbf{Z}$ denotes the mutual impedance matrix. The matrix $\mathbf{Z}$ is computed from the optimized SRA geometry at carrier frequency $f_c$ using the MATLAB Antenna Toolbox, and the corresponding effective channel is given by $\bar{\mathbf{H}}=\boldsymbol{\Gamma}_{\rm mc}\mathbf{H}_{\rm post}$. Following~\cite{Wong2024}, we consider dipoles with length $0.5\lambda$, width $0.005\lambda$, initial spacing $0.1\lambda$, and termination impedance $Z_T=50~\Omega$.

 \begin{figure}[ptb]
\centering
{\includegraphics  [height=2.53in, width=3.4in]{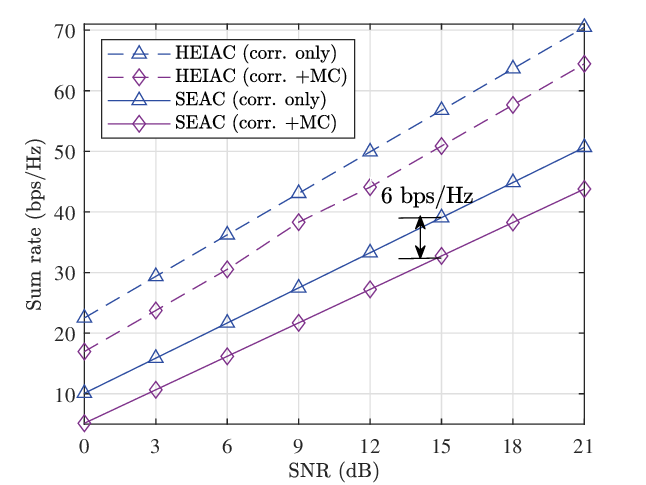}}
\caption{Effect of mutual coupling on the SEAC and HEIAC schemes.}
\color{black}
\label{mutual_coupling}
\end{figure}

In Fig.~\ref{mutual_coupling}, the impact of mutual coupling on the SEAC scheme $(M=2, S=4)$ and the HEIAC scheme $(M=2, S=4, N=2)$ is evaluated. For both schemes, the sum rate under the correlation-only (corr. only) model is consistently higher than that under the correlation-plus-mutual-coupling (corr. + MC) model, indicating that the mutual coupling effect is non-negligible in the considered compact array setup. For example, for SEAC at $\mathrm{SNR}=15$~dB, mutual coupling causes a performance loss of about $6$~bps/Hz. HEIAC consistently outperforms SEAC across both channel models, primarily due to the greater number of antennas in the considered configuration.

\section{Conclusion}
In this paper, we introduced a communication-oriented segment-wise SRA framework for correlation-aware antenna reconfiguration, where each soft robotic tentacle was divided into multiple independently controllable segments with elongation--retraction, bending, and sweeping motions. Based on this model, two antenna deployment schemes, namely SEAC and HEIAC, were proposed to enhance antenna-position controllability and reduce spatial correlation. In SEAC, antennas were deployed at segment endpoints and reconfigured by optimizing segment motion parameters, with the resulting sum-rate maximization problem solved via a PDD-PGA algorithm. In HEIAC, intra-segment MA-based reconfigurable antennas were incorporated to combine large-scale soft-robot deformation with fine-grained antenna positioning, and the joint optimization of segment deformation, intra-segment antenna positions, and antenna activation was solved using a BCD-PDD-PGA algorithm with greedy backward search. Simulation results showed that SEAC achieved 94.0\% and 37.9\% sum-rate gains over the 2D and 3D reconfigurable antenna baselines, and up to 49.3\% gain over the 3D baseline under compact array deployment. HEIAC also achieved 72.2\% and 32.1\% gains over the 2D and 3D baselines.

\end{document}